\documentclass[sigconf]{acmart}

\AtBeginDocument{%
  }

\usepackage{soul}  
\usepackage{color} 
\usepackage{xspace}
\usepackage{listings}
\usepackage{enumitem}
\newcommand{\eg}{{\it e.g.,\ }}

\newcommand{\ie}{{\it i.e.,\ }}

\usepackage{booktabs}
\definecolor{oxfordblue}{rgb}{0.0, 0.13, 0.28}
\definecolor{harvardcrimson}{rgb}{0.79, 0.0, 0.09}
\definecolor{dartmouthgreen}{rgb}{0.05, 0.5, 0.06}
\definecolor{princetonorange}{rgb}{1.0, 0.56, 0.0}
\definecolor{yaleblue}{rgb}{0.06, 0.3, 0.57}
\definecolor{usccardinal}{rgb}{0.6, 0.0, 0.0}
\definecolor{uclablue}{rgb}{0.33, 0.41, 0.58}
\definecolor{msugreen}{rgb}{0.09, 0.27, 0.23}
\definecolor{cornellred}{rgb}{0.7, 0.11, 0.11}
\definecolor{pomegranate}{RGB}{192, 57, 43}
\definecolor{anti-pomegranate}{RGB}{43,178,192}
\definecolor{alizarin}{RGB}{231, 76, 60}
\definecolor{anti-belize}{RGB}{185, 41, 56}
\definecolor{belize}{RGB}{41, 128, 185}
\definecolor{peter}{RGB}{52, 152, 219}
\definecolor{green}{RGB}{22, 160, 133}
\definecolor{anti-green}{RGB}{160,22,118}
\definecolor{turquoise}{RGB}{26, 188, 156}
\definecolor{pumpkin}{RGB}{211, 84, 0}
\definecolor{anti-pumpkin}{RGB}{0,22,211}
\definecolor{carrot}{RGB}{230, 126, 34}
\definecolor{wisteria}{RGB}{142, 68, 173}
\definecolor{anti-wisteria}{RGB}{99,173,68}
\definecolor{amethyst}{RGB}{155, 89, 182}
\definecolor{nephritis}{RGB}{39, 174, 96}
\definecolor{anti-nephritis}{RGB}{174,39,117}

\newcommand{\penguin}[1]{{#1}}
\newcommand{\pzh}[1]{{#1}}
\newcommand{\peng}[1]{{#1}}

\newcommand{\haoxiang}[1]{{#1}}
\newcommand{\yh}[1]{{#1}}
\newcommand{\fhx}[1]{{{#1}}}

\newcommand{\fanhx}[1]{{{#1}}}

\newcommand{\name}{{\textit{LitLinker}}}
    
\usepackage{booktabs}
\usepackage{multirow}
\usepackage{amsmath}
\usepackage{algorithm}
\usepackage{algorithmic}

\copyrightyear{2025}
\acmYear{2025}
\setcopyright{acmlicensed}\acmConference[CHI '25]{CHI Conference on Human Factors in Computing Systems}{April 26-May 1, 2025}{Yokohama, Japan}
\acmBooktitle{CHI Conference on Human Factors in Computing Systems (CHI '25), April 26-May 1, 2025, Yokohama, Japan}
\acmDOI{10.1145/3706598.3714111}
\acmISBN{979-8-4007-1394-1/25/04}

\begin{document}

\title{\name{}: Supporting the Ideation of Interdisciplinary Contexts with Large Language Models for Teaching  Literature in Elementary Schools}

\author{Haoxiang Fan}
\email{fanhx6@mail2.sysu.edu.cn}
\orcid{0009-0000-5729-8491}
\affiliation{%
  \institution{Sun Yat-sen University}
  \city{Zhuhai}
  \country{China}
}

\author{Changshuang Zhou}
\authornote{Both authors contributed equally to this research.}
\email{mc34188@um.edu.mo}
\affiliation{%
  \institution{University of Macau}
  \city{Macau}
  \country{Macao}
}

\author{Hao Yu}
\authornotemark[1]
\email{yuhao53@mail2.sysu.edu.cn}
\affiliation{%
  \institution{Sun Yat-sen University}
  \city{Zhuhai}
  \country{China}
}

\author{Xueyang Wu}
\email{xwuba@connect.ust.hk}
\affiliation{%
  \institution{NeurlStar}
  \city{Shenzhen}
  \country{China}
}

\author{Jiangyu Gu}
\email{3299158551@qq.com}
\affiliation{%
  \institution{Xiangzhou Experimental School of Zhuhai}
  \city{Zhuhai}
  \country{China}
}

\author{Zhenhui Peng}
\authornote{Corresponding author.}
\email{pengzhh29@mail.sysu.edu.cn}
\orcid{0000-0002-5700-3136}
\affiliation{%
  \institution{Sun Yat-sen University}
  \city{Zhuhai}
  \country{China}
}

\renewcommand{\shortauthors}{Fan et al.}
\renewcommand{\shorttitle}{\name{}}

\begin{abstract}

\pzh{
Teaching literature under interdisciplinary (\eg science, art) contexts that connect reading materials has become popular in elementary schools. However, constructing such contexts is challenging as it requires teachers to explore substantial amounts of interdisciplinary content and link it to the reading materials. In this paper, we develop \name{} via an iterative design process involving 13 teachers to facilitate the ideation of interdisciplinary contexts for teaching literature. Powered by a large language model (LLM), \name{} can recommend interdisciplinary topics and contextualize them with literary elements (\eg paragraphs, viewpoints) in the reading materials. A within-subjects study (N=16) shows that compared to an LLM chatbot, \name{} can improve the integration depth of different subjects and reduce workload in this ideation task. Expert interviews (N=9) also demonstrate \name{}'s usefulness for supporting the ideation of interdisciplinary contexts for teaching literature. We conclude with concerns and design considerations for supporting interdisciplinary teaching with LLMs.  
}

\end{abstract}




\maketitle
\section{Introduction}

The integration of knowledge from different disciplines within literature instruction at the elementary school level 
has been proven to enhance student learning outcomes~\cite{navrotskaya2024reading, lindvig2019different}. 
Contextualization, as a pedagogical approach, makes the curriculum more meaningful and practical for students, which is essential for effective learning~\cite{fernandes2013curricular}. Learning in an interdisciplinary context not only enhances students' comprehensive learning abilities but also increases their motivation to learn. 
However, the ideation of a suitable interdisciplinary context \pzh{is challenging}. 
Firstly, human knowledge is categorized into distinct disciplines, \peng{and people with an occupation usually focus their expertise on a single field~\cite{palmer2010information}.} 
The development of an interdisciplinary context requires literature teachers to engage in the continuous exploration and evaluation of substantial amounts of information from unfamiliar fields (\eg mathematics, science, and art).
Secondly, elementary school teachers frequently face constraints related to limited time outside of the classroom~\cite{dias2017challenges}, which do not allow them to conduct comprehensive searches for interdisciplinary knowledge. 

Recent advances in large language models (LLMs) show great potential to address the challenges faced by teachers in ideating interdisciplinary contexts.
Such as GPT-4~\footnote{\url{https://openai.com/index/gpt-4}} and GLM-4~\footnote{\url{https://github.com/THUDM/GLM-4}}, possess extensive knowledge across various domains, owing to their pre-training on large text datasets~\cite{NEURIPS2020_1457c0d6}.
\fhx{
Also, LLMs' capabilities in long-context understanding~\cite{naveed2023comprehensive} enable them to synthesize information from extensive textual materials effectively. 
}
Within the field of human-computer interaction (HCI), the LLM-empowered tools have emerged to facilitate interdisciplinary information exploration~\cite{zheng2024disciplink}, lesson plan preparation based on educational theories~\cite{fan2024lessonplanner}, and support learners in critical and creative thinking~\cite{yuan2023critrainer, shaer2024ai}. 
However, neither conversational applications of LLMs (\eg ChatGPT) nor these LLM-powered tools can offer a comprehensive solution for the challenges at hand. 
Firstly, the process of ideating interdisciplinary contexts from a variety of reading materials necessitates that educators engage in a thorough exploration of both the context and the materials. 
This engagement enables teachers to discern the connections between the context and the reading materials, rather than reading the responses generated by LLMs. Without such an in-depth exploration, fostering a comprehensive understanding of the context and materials becomes challenging, which is essential for effective teaching. 
Secondly, the suggested interdisciplinary contexts need to be effectively integrated into the classroom environment.
However, due to a limited understanding of 
\haoxiang{teaching practices} and the cognitive backgrounds of elementary students, LLM-generated content often fails to align with the cognitive and pedagogical demands of elementary education~\cite{fan2024lessonplanner}.
Consequently, the generated content may require adjustments by literature teachers or may even be unusable. 
\haoxiang{
Lastly, educators, particularly those who are not so familiar with LLMs, have to allocate significant time and effort in constructing complex prompts and understanding the intricate outputs generated by these systems.
In summary, there is a need for an interactive tool that assists educators in effectively exploring and generating ideas for interdisciplinary contexts within the classroom.
}

\penguin{In this paper, 
\fanhx{we focus on the teaching scenarios of Chinese language courses in China and}
introduce \name{}~\footnote{\fanhx{It is open-sourced and available on \url{https://github.com/fanhaoxiang1/LitLinker}}}, 
an LLM-powered interactive system that supports teachers in ideating diverse interdisciplinary contexts for teaching literature in elementary schools. 
}
\penguin{To develop \name{}, we follow a user-centered design approach that involves in total 13 Chinese literature teachers in three interview sessions, prototyping, and evaluation.}
\penguin{\name{} is} 
a web application powered by the LLM GLM-4. When users select relevant disciplines 
and reading materials for exploration, \name{} initially recommends interdisciplinary topics 
and corresponding analyses based on the themes and concepts of the selected materials.
\haoxiang{
Subsequently, \name{} facilitates the exploration of relationships between various literary elements (\eg paragraphs, sentences, and viewpoints within reading materials) and the selected topics.
}
Users can bookmark topics they focused on and ask the LLM any questions related to the topics and reading materials. 
Finally, \name{} produces a \penguin{lesson plan that includes a} course outline, an introductory overview, and recommended classroom activities.


We evaluate \name{} through two studies
\penguin{that answer three research questions (RQs), \ie how would \name{} affect the interdisciplinary context exploration RQ1) outcomes and RQ2) process, and RQ3) how would users perceive the usability and creative support of \name{}}. 
\penguin{Experiment I is a within-subjects design that quantitatively answers these three RQs with 16 novices of Chinese literature teaching, while Experiment II is a qualitative study that gains answers to the RQs from 10 elementary Chinese literature teachers with varying teaching experience.
The quantitative results indicate that \name{} leads to deeper integration of different subjects in the outcome lesson plan, is deemed significantly more satisfying and efficient, and significantly reduces task workload. 
}
The qualitative analysis reveals high satisfaction with the generated results and a strong appreciation for \name{} in facilitating teachers to develop interdisciplinary courses.

In summary, this paper has three main contributions. 
First, \penguin{we introduce \name{}, 
an interactive system that supports the ideation of interdisciplinary contexts for literature classes.} 
Second, through a within-subjects study and expert interviews, we provide empirical evidence of \name{}'s effectiveness and usefulness. 
Third, based on our design process and findings, we offer design implications for future LLM-based systems that assist teachers. 

\penguin{
In the rest of this paper, we first review literature that inspires our work (\autoref{sec:related_work}), followed by the design process (\autoref{sec:design_process}) and implementation (\autoref{sec:system_design}) of \name{}.
Then, we present two experiments (\autoref{sec:experiment_1}, \autoref{sec:experiment_2}) that evaluates \name{}. 
We finally discuss the implications of our work to HCI and education communities (\autoref{sec:discussion}). 
} 
\section{Related Work}
\label{sec:related_work}
\subsection{Interdisciplinary and Literature Instruction}
Interdisciplinary education equips students to think across subject boundaries, addressing complex global challenges and enhancing critical thinking, problem-solving, and collaboration skills~\cite{rhoten2006interdisciplinary}.
Its growing importance is evident in fostering adaptable, innovative learners for a rapidly changing world~\cite{you2017teach}. 
By applying language in real-world contexts, disciplinary approaches promote engagement~\cite{towhidnejad2011introducing}. Interdisciplinary education offers key advantages in K-12 language learning, promoting critical thinking, creativity, and motivation by connecting learning to real-world issues~\cite{bestelmeyer2015collaboration, hardre2013teachers}. 
Empirical studies often focus on STEM fields, such as~\citet{english2016stem}’s integration approach and~\citet{sarama2017interdisciplinary}'s C4L curriculum, which combines language learning with early childhood development. Also, \citet{czerniak2014interdisciplinary} demonstrated how interdisciplinary science teaching enhances language development.

With the introduction of the new curriculum standards by China's Ministry of Education in 2022, interdisciplinary curriculum design was formally included in the ``expansive learning task group'' of Chinese language education, explicitly outlining requirements for interdisciplinary teaching~\footnote{\url{http://www.moe.gov.cn/srcsite/A26/s8001/202204/W020220420582344386456.pdf}}. 
Guided by these new standards, interdisciplinary teaching in primary school Chinese language classes can bring a fresh perspective to Chinese language education. By the design of diverse interdisciplinary activities, course content is integrated to cultivate students' active learning and inquiry skills. 
Additionally, reading instruction fosters students' interdisciplinary thinking abilities, promotes cross-subject integration, and stimulates students' imagination. 
\pzh{However,}
few works focus on integrating literature with other subjects and customizing technologies for diverse teaching environments. 
\peng{Teachers could face challenges in preparing for interdisciplinary literature teaching due to inadequate professional development}~\cite{boix2010interdisciplinarity, yarker2012analysis}. 
\peng{Our work is motivated by the trend and benefits of interdisciplinary literature teaching and aims to help teachers ideate interdisciplinary contexts for teaching literature in elementary schools.} 

\subsection{Interactive Systems that Support Brainstorming and Ideation}

Interactive systems have been widely used to facilitate the creative thinking processes, including art design~\cite{heyrani2021creativegan, frich2018hci}, writing~\cite{gero2022sparks}, and learning~\cite{jin2024codetree}. 
Creativity Support Tools (CSTs) provide numerous advantages due to their interactive nature. 
For instance, C2Ideas~\cite{hou2024c2ideas}, an innovative system for designers to ideate color schemes, provides users with an interactive workflow that aligns with traditional interior design methods to facilitate the design process. 
Users are required to input their initial design intentions and customize the intermediate results, after which the system generates good outcomes. Furthermore, in the field of supporting creative activities within interdisciplinary environments, DiscipLink~\cite{zheng2024disciplink} assists users in making sense of information by generating exploratory questions, expanding queries, and extracting themes and connections among academic papers. 
Users can freely explore the carefully designed Orientation View, Exploration View, and Collection View of the user interface, a layout that we also use.

More and more creativity support interactive systems are being designed for specific tasks in particular professions~\cite{louie2020novice, petridis2023anglekindling, choi2024creativeconnect, li2024diaryhelper}. 
For example, AngleKindling~\cite{petridis2023anglekindling} is designed to assist journalists in exploring diverse angles for reporting on press releases. 
This work is similar to ours, as the creativity support also comes from analyzing large amounts of textual material, and our work also leverages the capabilities of LLMs. 
There also has been a growing trend toward the integration of human-computer interaction (HCI) and education. 
This integration has led to the development of interactive systems aimed at enhancing educational outcomes by promoting users' creative thinking~\cite{zhang2022storydrawer}. NaCava~\cite{yan2023nacanva} is a mobile interactive system designed to facilitate nature-inspired creativity for children. 
It enhances children's multidimensional observation and engagement with nature by encouraging the collection of multi-modal materials and using them in a creation process.

However, few studies focus on developing creativity support tools for teachers to help them design classroom activities to promote student learning, especially in interdisciplinary setups. In our work, we create \name{} aimed at teachers to help them ideate interdisciplinary contexts in education, offering creativity support to facilitate the ideation process.

\subsection{Large Language Models in Education}
Large Language Models (LLMs) are increasingly utilized across various educational domains due to their capacity to facilitate personalized learning experiences and enhance accessibility for all learners~\cite{mogavi2024chatgpt}. 
They support self-learning to enhance personal skills in diverse areas, including programming~\cite{yilmaz2023augmented}, problem-solving ability~\cite{kasneci2023chatgpt}, and writing skills~\cite{shidiq2023use}. Additionally, they also support various activities within classroom environments, such as helping with homework, reviewing~\cite {mogavi2024chatgpt}, and assisting students in project-based learning~\cite{zheng2024charting}. 
Specifically, within the K-12 educational domain, LLMs have demonstrated their effectiveness in providing assistance across multiple subjects, including foreign languages, programming, and mathematics~\cite{mogavi2023exploring}. This capability is attributed to their pre-training on large datasets across various fields~\cite{NEURIPS2020_1457c0d6}.
\penguin{
Recent work also suggested that designing multiple LLM agents to simulate human collaboration can enhance the quality of the outputs on domain-specific tasks, such as translation~\cite{wu2024transagents} and software development~\cite{du2024multi}.
Inspired by these works, we leverage the potential of LLMs and deploy them as agents to generate precise and comprehensive interdisciplinary information in our study.
}

LLMs also promote instructional activities for teachers. 
For instance, RetLLM-E~\cite{mitra2024retllm} provides instructional support by delivering context-aware, high-quality answers to student questions using LLMs. This research illustrates that retrieved context can significantly increase the quality of LLM-generated responses, thereby informing the advantages of 
utilizing retrieval-augmented generation~\cite{lewis2020retrieval} for precise instruction. 
Furthermore, LessonPlanner~\cite{fan2024lessonplanner}, an LLM-driven tool that formulates structured lesson plans based on educational theories, has been empirically validated to assist teachers in increasing the efficiency of lesson preparation and the quality of lesson plans. 
The objectives of our work are similar to theirs, as both emphasize the quality of outcomes and the user experience.

Previous studies have demonstrated the potential of LLMs in education, but few works focus on teachers' preparation for interdisciplinary instruction scenarios.
In our work, we conduct an iterative design with 13 teachers to understand their views on how LLMs can facilitate interdisciplinary literature instruction and support them.
\pzh{
\section{Design Process and Principles of \name{}}
\label{sec:design_process}
Our work aims to support elementary literature teachers in effectively identifying suitable interdisciplinary contexts for their instructions, which can be used in their later lesson plans and classroom activity designs. 
\penguin{
Our design process and evaluation of \name{} involve in total of 17 Chinese language teachers (\autoref{tab:teachers}) in an elementary school in mainland China. 
Specifically, in the design process, we involved E1-E7 in the foundational study and I1-I6 in the evaluation of the prototype. 
In the evaluation of \name{} with teachers (\ie Experiment II), we involved E1-E5 again and E8-11. 
In Experiment II, I1-I6 also contributed findings about the unchanged features between \name{} and its prototype. 
}



\begin{table*}[htbp]
\caption{17 Chinese language teachers participated in the iterative design process and expert interviews (\ie Experiment II). Among them, there were 6 males and 11 females, with teaching experience ranging from 3 to 29 years. Two participants did not provide information on their teaching experience. This table also includes their experience in reading projects and interdisciplinary projects.}
\Description{17 Chinese language teachers participated in the iterative design process and expert interviews (\ie Experiment II). Among them, there were 6 males and 11 females, with teaching experience ranging from 3 to 29 years. Two participants did not provide information on their teaching experience. This table also includes their experience in reading projects and interdisciplinary projects.}
\label{tab:teachers}
\begin{tabular}{@{}cccccc@{}}
\toprule
\textbf{Involvement}                                                                                                                               & \textbf{ID} & \textbf{Gender} & \begin{tabular}[c]{@{}c@{}}\textbf{Teaching Experience} \cr \textbf{(years)}\end{tabular} & 
\begin{tabular}[c]{@{}c@{}}\textbf{Participation in} \cr \textbf{Reading Projects}\end{tabular} & 
\begin{tabular}[c]{@{}c@{}}\textbf{Participation in} \cr \textbf{Interdisciplinary Projects}\end{tabular} \cr \midrule
\multirow{5}{*}{\begin{tabular}[c]{@{}c@{}}\textit{Foundational}\cr \textit{Study}\cr \textit{\&}\cr \textit{Experiment II}\end{tabular}}                                        & E1          & F               & 5                                                                               & Y                                                                                     & Y                                                                                               \cr
                                                                                                                                                   & E2          & M               & 7                                                                               & Y                                                                                     & Y                                                                                               \cr
                                                                                                                                                   & E3          & F               & 14                                                                              & Y                                                                                     & Y                                                                                               \cr
                                                                                                                                                   & E4          & M               & 6                                                                               & Y                                                                                     & Y                                                                                               \cr
                                                                                                                                                   & E5          & M               & 4                                                                               & Y                                                                                     & Y                                                                                               \cr \hline
\multirow{2}{*}{\begin{tabular}[c]{@{}c@{}}\textit{Foundationall}\cr \textit{Study}\end{tabular}}                                                             & E6          & F               & -                                                                               & Y                                                                                     & Y                                                                                               \cr
                                                                                                                                                   & E7          & F               & -                                                                               & Y                                                                                     & Y                                                                                               \cr \hline
\multirow{6}{*}{\begin{tabular}[c]{@{}c@{}} \textit{\penguin{Evaluation of}}\cr \textit{\penguin{Prototype (re-usable}}\cr \textit{\penguin{findings are presented}}\cr \textit{\penguin{in Experiment II)}}\end{tabular}} & I1          & F               & 27                                                                              & Y                                                                                     & Y                                                                                               \cr
                                                                                                                                                   & I2          & F               & 8                                                                               & Y                                                                                     & Y                                                                                               \cr
                                                                                                                                                   & I3          & F               & 5                                                                               & N                                                                                     & N                                                                                               \cr
                                                                                                                                                   & I4          & M               & 11                                                                              & Y                                                                                     & Y                                                                                               \cr
                                                                                                                                                   & I5          & F               & 5                                                                               & Y                                                                                     & N                                                                                               \cr
                                                                                                                                                   & I6          & M               & 5                                                                               & Y                                                                                     & N                                                                                               \cr \hline
\multirow{4}{*}{\begin{tabular}[c]{@{}c@{}}\textit{\penguin{Expert}}\cr \textit{\penguin{Interviews}}\end{tabular}}                                                              & E8          & M               & 3                                                                               & N                                                                                     & N                                                                                               \cr
                                                                                                                                                   & E9          & F               & 6                                                                               & Y                                                                                     & N                                                                                               \cr
                                                                                                                                                   & E10         & F               & 29                                                                              & N                                                                                     & Y                                                                                               \cr
                                                                                                                                                   & E11         & F               & 17                                                                              & Y                                                                                     & N                                                                                               \cr \bottomrule
\end{tabular}
\end{table*}

\subsection{Design Process}
\penguin{
We generally followed a user-centered approach to plan our design process. 
First, to understand users' practices and involve them in the design of \name{}, we conducted three sessions of semi-structured interviews with an experienced literature teacher E1, who led a seven-member (E1 - E7) interdisciplinary literature course design team that indirectly contributed to the interviews. 
We were not able to have direct discussions with E2 - E7 due to their inconvenience during the semester. 
Then, we developed a workable prototype of \name{} and evaluated it with another six teachers (I1 - I6).  
We gathered their insights to inform our design goals for \name{} presented in this paper. 
}
\subsubsection{Foundational Study}
We closely worked with E1 to identify the practices, challenges and needs for support of ideating interdisciplinary contexts for teaching literature in elementary schools. 
Over the past two years, E1 has spearheaded a team of seven individuals (E1-E7) in the exploration and implementation of interdisciplinary literature instruction within elementary school Chinese courses. 
\fanhx{
To gain a comprehensive understanding of user needs, we progressively conducted three sessions of semi-structured interviews with E1 in April, June, and July 2024, lasting 38 minutes, 53 minutes, and 45 minutes, respectively.
}
Before each session, we communicated the purpose of the interview to E1 and requested that she engage with her team to compile records of their meetings for discussing the topics in the intended interview. 
We documented each interview session with E1 through audio and video recordings. 

In Session 1, 
\fanhx{we asked E1 to present their current practices of teaching literature in interdisciplinary contexts, with previously developed lesson plans and assignments in her team.
} 
The discussion also focused on the potential of \fanhx{AI (\eg what do you think AI can support you (in your lesson planning in previous))}, and an interactive system to facilitate the design of interdisciplinary literature contexts, including the integration of art and history into the assignments \peng{of literature reading}. 
\peng{After this section}, two authors brainstormed potential features of a system for supporting the ideation of interdisciplinary contexts and sent E1 a document that explains these features. 
We requested E1 to engage in a discussion with her group members to identify any additional or incorrect points about the \peng{potential system}. 
In Session 2, E1 \peng{came back with positive feedback from her team on each potential feature}.
\fanhx{We asked her to further explain their general process for designing interdisciplinary contexts as a team, emphasizing the distinct responsibilities and cognitive processes of each teacher involved.} 
Additionally, she presented the proposed interaction model. 
After this session, E1 had a group meeting with her team and came up with a template that defines the anticipated outcomes of our system. 
In Session 3, we introduced how a system works utilizing LLM agents \peng{to simulate roles in a team for ideating the interdisciplinary contexts, as suggested by E1 in Session 2}. 
We presented two example outcomes produced using our predefined prompts and intermediate outputs 
\fanhx{to ask for her opinions on these prompts and outputs (\eg whether these intermediate outputs were helpful? If the prompts aligned well with your thoughts?).
}
}

\pzh{
\subsubsection{Development and Evaluation of \name{} Prototype}
\peng{
After Session 3, two of the authors utilized the thematic analysis method to analyze the transcribed recordings and all textual content derived from the foundational study. 
The analysis yielded four summarized Design Principles as described in the following \autoref{sec:principles}. 
We then worked on the implementation of a workable prototype that chains different LLM agents in a structured process to help teachers think of interdisciplinary contexts for teaching the literature materials in the textbooks. 
}
We evaluated our workable prototype with another six elementary school Chinese language teachers (I1 - I6, 3 Male, 3 Female), as shown in \autoref{tab:teachers}. 
Each evaluation lasted approximately 30 to 45 minutes and comprised four parts: (1) an introduction to the background, which included the concepts of interdisciplinary literature instruction, and the theory of contexts of instruction; (2) a brief tutorial on the prototype; (3) a think-aloud study in which participants freely explored the prototype and spoke out their thoughts; and (4) a semi-structured interview \peng{for their comments and suggestions on the prototype}. With the participants' consent, we conducted the evaluation offline and recorded audio and video.

At this stage, we assessed the prototype's usability (\eg whether different functions were well-integrated), user perception (\eg user interaction with the prototype and any additional cognitive load), and the quality of the system's outcomes. Furthermore, we collected suggestions for improving the prototype, particularly regarding user interface design and additional functionalities. 
\penguin{These feedback and suggestions are presented in \autoref{sec:formative_findings}, which inform the design principles (\autoref{sec:principles}) of \name{}.}
For the six teachers' (I1-6) feedback on the same features in the prototype and final version of \name{}, we incorporate it in the results of Expert Interviews in \autoref{sec:experiment_2}. 
}


\penguin{
\subsection{Findings}
\label{sec:formative_findings}
}
\fhx{
Two of the authors utilized thematic analysis~\cite{braun2012thematic} to inductively code and summarize the \penguin{practices, challenges, requirements, and concerns} from transcribed recordings \penguin{in the design process. The key themes are shown below.} 
\fanhx{One author first iteratively coded the data, while the other carefully reviewed the codes to ensure accuracy. After discussions, they reached a consensus and identified six primary themes. These findings are shown below.}
}

\penguin{
\textbf{Finding 1: In practice, teachers usually engage in reverse thinking when ideating interdisciplinary context}.}
\penguin{Our teachers mentioned two intellectual paradigms.}
One paradigm referred to as ``forward thinking'', resembles ``deductive reasoning'', in which
teachers create abstract connections from a limited number of reading materials (typically 3-5 texts) and develop a concrete and reasonable context. 
The other paradigm, which is termed ``reverse thinking'' and analogous to ``inductive reasoning'', is a more habitual cognitive process employed by teachers. 
Teachers would like to first select interdisciplinary contexts that they deem suitable and then identify appropriate texts from a broader text pool, after which they refine the connections between the identified text and the context. 
\textit{``For us, a good context often arises from a sudden inspiration, which we then backtrack to complete the ideation of what texts can connect to this context and how''} (E1).

\penguin{
\textbf{Finding 2: It is challenging to identify the connections between the established context and the reading materials  in the process of ``reverse thinking''.} 
}
Teachers must evaluate the effectiveness of the connections in enhancing students' understanding of both literature and its associated subjects, as well as in stimulating their interest. 
\textit{``It is quite difficult and usually takes a long time for our team to ensure that based on the reading materials, our teaching activities connected by the context can indeed help students gain knowledge''} (E1).

\penguin{
\textbf{Finding 3: Teachers require support at three levels of granularity when analyzing the reading materials and contexts.}
}
The first level is in-depth single-text analysis, which \textit{``explains how the elements of a given article relate to the context''} (E1). 
The second level is pairwise comparison, where comparative reading has been demonstrated to be an effective method for understanding texts, \textit{``allowing articles to `disappear in pairs' by analyzing the similarities and differences in relation to the context''} (E1). 
The third level is multi-text-driven exploration, which necessitates that the system should support the comprehensive linking of all texts selected by the teacher. Therefore, this level requires \name{} to conduct a thorough deconstruction of contexts, extract meaningful connections, and convey these connections to the teachers. 

\penguin{
\textbf{Finding 4: 
Teachers require detailed instructional activities based on the selected contexts.}}
\penguin{As E1 summarized after the meeting with her teaching team, the outcome plan of several lessons surrounding a context should include targeted reading materials and analysis in each lesson, an introduction facilitating students' engagement, and related teaching activities.}
Also, in the evaluation study of \name{} prototype, three teachers (I1, I2, I5) indicated that the system outputs should be more detailed and reduce human effort in modifying them for the later concrete plans for each lesson. 
\textit{``The overall structure of the output is good, but I hope it can be more detailed; for example, providing more in-class and extracurricular activities related to the theme, so we can use them directly''} (I1). Therefore, we incorporated recommendations for literature and interdisciplinary course activities in the refined \name{}.

\penguin{
\textbf{Finding 5: Teachers are concerned the quality and reliability of the content purely generated by LLMs.}
}
In the evaluation study with six teachers (I1-6), our prototype generated traditional subject-related contexts using the LLM with specific templates and cognitive backgrounds of elementary students, without fine-tuning or retrieval-augmented generation (RAG). 
Three teachers (I4, I5, I6) expressed concerns about the quality of the LLM-generated content. \textit{``The content generated for the art subject is quite repetitive''} (I5). \textit{``We need to establish a dedicated article database for science as well, since many of our articles are highly relevant to science''} (I4). 

\penguin{
\textbf{Finding 6: Teachers suggest six metrics for evaluating the outcome of interdisciplinary literature lesson plan.}
As established by the team of E1-E7, the metrics are: 
}
\begin{itemize}
    \item \textbf{Appropriateness of Context}
    \begin{itemize}
        \item \textit{Content Alignment:} Does the context accurately cover the content of the selected materials?
        \item \textit{Internal Logic:} Is there a logical connection between the context and the selected materials?
    \end{itemize}
    \item \textbf{Alignment with Educational Objectives}
    \begin{itemize}
        \item \textit{Curriculum Standards:} Does the content comply with national curriculum standards and teaching guidelines?
        \item \textit{Subject Goals:} Does it help achieve specific goals of language education, such as reading comprehension and writing skills?
    \end{itemize}
    \item \textbf{Depth of Integration}
    \begin{itemize}
        \item \textit{Subject Integration:} Does it effectively integrate knowledge from different subjects?
        \item \textit{Knowledge Transfer:} Does it promote the application of language arts knowledge in other subject contexts?
    \end{itemize}
\end{itemize}



\subsection{Design Principles} \label{sec:principles}

\penguin{Based on the findings from our design process and related literature, we derive four design principles of \name{}.}

\penguin{
\textbf{DP1: \name{} should provide step-by-step support that aligns with teachers' habitual practices in interdisciplinary context ideation}.
Tailoring the assistance to users' habitual practices (\eg active students' behaviors or teachers' behaviors) is a commonly enacted principle in previous interactive systems in educational scenarios~\cite{fok2024qlarify,liu2024classmeta,fan2024lessonplanner}. 
In the task of interdisciplinary context ideation, as revealed in Finding 1, \name{} should support step-by-step context ideation through reverse thinking, which is a habitual practice of our teachers. 
\fanhx{In this practice, as E1 shared,} teachers take various roles to analyze the potential contexts 
\fanhx{(we note this role as Context Analyst)}, 
analyze the texts in the reading materials, connect them to the contexts, 
\fanhx{and discuss the approaches (Text Analyst and Text Reviewer)}. 
After that, teachers try to summarize the contexts and associated reading materials into an actionable lesson plan \fanhx{(Context Summarizer)}. 
\name{} can prompt LLMs to play different roles when supporting teachers in each of these steps. 
We do not chase for generating one-step context ideation outcome with one LLM prompt, because teachers desire necessary human input in each step, and enabling multiple LLM agents to simulate human-human collaboration has been proven to improve output quality~\cite{wu2024transagents, du2024multi}. 
}
\penguin{
\textbf{DP2: \name{} should provide teachers with detailed analyses of the contexts, reading materials, and their relationships.}
Our teachers reported that it was challenging to identify the connections between contexts and reading materials (Finding 2) and desired support during the analyses (Finding 3). 
Previous HCI works have demonstrated the strengths of LLMs in analyzing and connecting complex information~\cite{zheng2024disciplink, chi24_Selenite}. 
Similarly, to satisfy user requirements (Finding 3) in our task, \name{} can leverage LLMs to recommend contexts, explain them in detail, identify relevant texts in the reading materials, and assess the relationship between the contexts and texts. 
}

\penguin{
\textbf{DP3: \name{} should document the ideation outcomes in a lesson plan that aligns with the established educational practices in interdisciplinary literature teaching.}
A teacher without a lesson plan may struggle to effectively deliver the knowledge and objectives of the lesson~\cite{iqbal2021rethinking}. 
Finding 4 suggests that the lesson plan should contain detailed instructional activities, including the related reading materials and in-class activities, based on the selected contexts. 
To make it further aligned with educational practices, \name{} can adopt the six evaluation metrics of the outcome lesson plan (Finding 6) to guide the generation of ideation outcomes. 
}

\pzh{

\penguin{
\textbf{DP4: \name{} should include database of interdisciplinary contexts and reading materials and provide flexible user control to achieve high-quality ideation outcomes.} 
Prior research on the impact of LLMs in primary education indicates that generating false content is a disadvantage that may lead to ``information pollution'' for children~\cite{adeshola2023opportunities, murgia2023chatgpt}.
Finding 5 also indicates that LLMs sometimes were unable to create content that meets teachers' needs when lacking access to educational resources. 
To generate high-quality outcomes, as inspired by previous works~\cite{yazici2024gelex, khanal2024fathomgpt}, 
\name{} could ground the content generation on diverse real-world contexts and reading materials. 
\name{} should also support teachers to freely edit and question any content (\eg texts in reading materials, outcome lesson plan) to make sure that they understand the content they are going to use in literature teaching. 
}

}
\pzh{
\section{Design of \name{}}
\label{sec:system_design}
Based on the identified design principles, we introduce \name{}, a human-AI collaborative system to support elementary school literature teachers in ideating interdisciplinary contexts.
\name{} not only facilitates teachers' ability to explore various contexts and reading materials but also delivers structured outputs derived from this exploration process and the teachers' thought processes. 
Inspired by prior research~\cite{zheng2024disciplink} and tailored to our interdisciplinary context exploration scenario, the primary interface of \name{} consists of three parts: Contexts Exploration View, Texts Exploration View, and Collection View. 
After confirming the subjects and importing the reading materials, the system aligns with the user's habitual thinking process by selecting or specifying potentially related interdisciplinary contexts within Context Exploration View \penguin{(DP1)}. 
Based on the selected context, the system enables multi-text-driven exploration (DP2). 
\penguin{Specifically,} \name{} extracts connections between the context and literacy elements, displaying specific details in either Context Exploration View or Texts Exploration View, decided upon whether they are context-oriented or text-oriented. 
In Texts Exploration View, users can conduct in-depth analyses of individual texts and perform pairwise comparisons of any selected texts. 
\peng{In these two exploration processes}, users have the opportunity to chat with the LLM at any time and modify the analyses provided by the system \penguin{(DP4)}. 
Users can add analyzed themes and corresponding texts into the collection, allowing the system to generate structured ideation outcomes based on their selections (DP3). 
In \name{}, all generated content relies on a context database, \peng{prompt} templates, and generation guidelines (DP3, DP4). 
In the following subsections, we will describe how users can interact with \name{} and the detailed design and implementation of \name{}.
}

\begin{figure*}[ht]
  \centering
  \includegraphics[width=\linewidth]{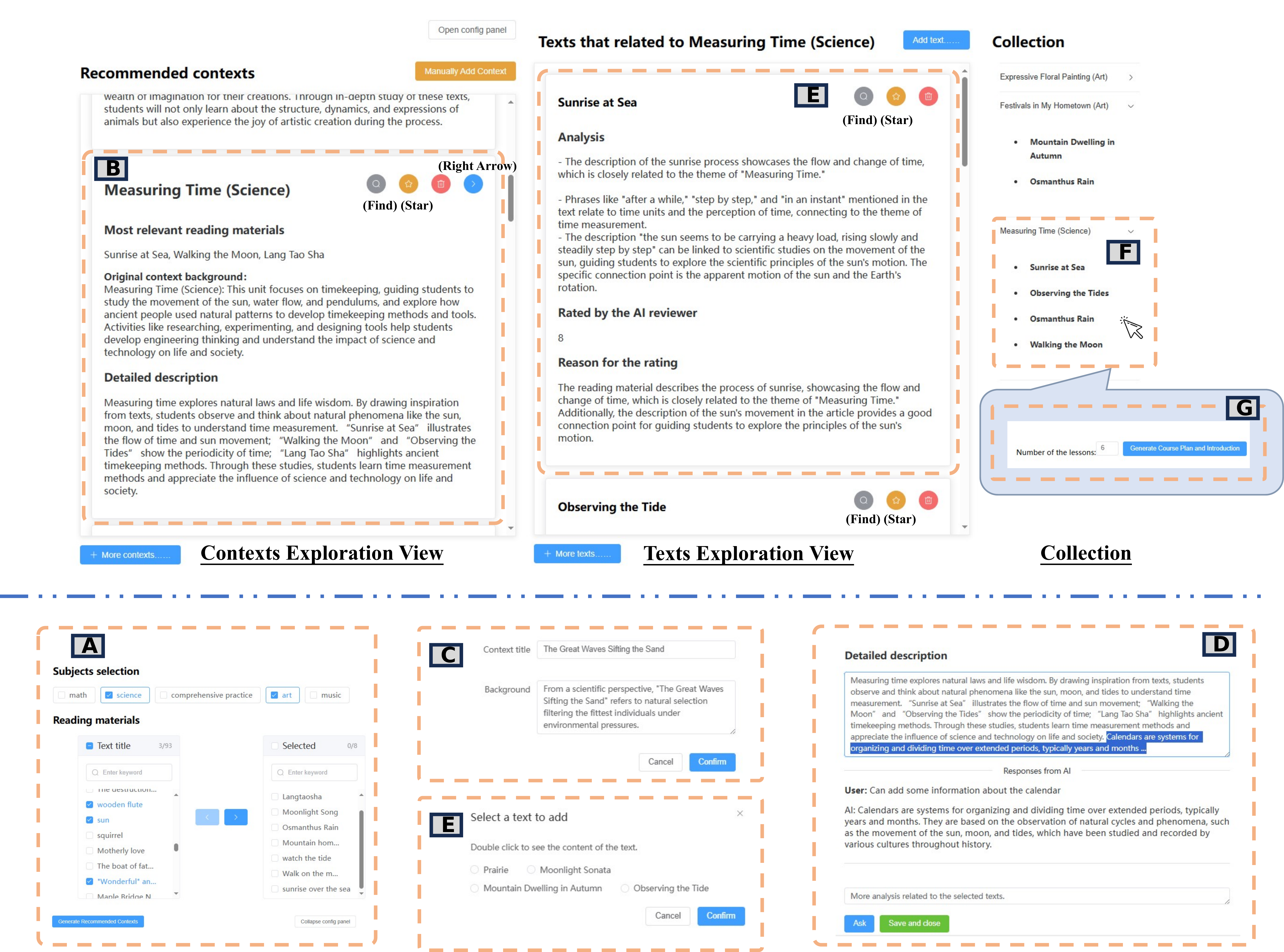}
  \caption{The user interface of \name{}, translated from Chinese either by Google Translator or manually.}
  \Description{The user interface of \name{}, translated from Chinese either by Google Translator or manually.}
\label{fig:system_interface}
\end{figure*}

\subsection{\name{} Interface and Interaction Design}
\pzh{To initiate}
the exploration process, users are required to select the interdisciplinary subjects and reading materials they wish to explore \pzh{(\autoref{fig:system_interface} A)}.
By clicking the ``Collapse Config Panel'' button (~\autoref{fig:system_interface} A turned into ``Open config panel'' at the top of the page after clicking on), users confirm their settings, which subsequently expands Contexts Exploration View. Upon clicking the ``Generate Recommended Contexts'' \pzh{button, \name{}} displays a series of context cards (B), each containing the context title, the titles of the most relevant reading materials, the original context from Context Pool (see~\autoref{sbsbsc:contextpool}), and a detailed description of the context.
Users may click the ``Find'' button on any context card of interest to further explore and edit the description of the context, allowing them to ask the LLM any questions  \peng{(D)} related to the theme and its connections (\eg relate this context to additional musical works?) to aid in understanding and exploration. The description becomes editable, enabling users to reconstruct it based on the LLM's responses. Users can also click ``Star'' to favorite a context and click the ``Right Arrow'' to specify the theme associated with texts in Texts Exploration View  \pzh{(B)}. 
If users are dissatisfied with a context, they have the option to click the ``Delete'' button to remove it. Users can click ``Manually Add Context'' to provide the title and background of a context in a pop-up window \pzh{(C)}, allowing \name{} to tailor a suitable context based on the information provided.

The middle part of the page is designated as Texts Exploration View (\autoref{fig:system_interface}), which facilitates a comprehensive examination of the interrelationships among various reading materials and specified contexts. 
Texts that are most relevant to the selected context are displayed at the top in Texts Exploration View, with each reading material accompanied by the analysis.
This analysis encompasses the relationships among sentences, paragraphs, and the context (\ie in-depth single-text analysis). 
To further support user exploration, inspired by~\cite{wu2024transagents} and the workflow of the teachers, additional ratings and critical recommendations are provided for LLM-generated analysis.
Similar to Contexts View, users have the option to click the ``Find'' button on any card (E) to ask the LLM any questions about the contexts related to the reading material (\eg Can you elaborate on the relationship between the descriptions of scenery in the text and the context?), and users can also modify the analysis. 
Users can click ``Star'' to favorite a reading material under the relevant context or click ``Delete'' if they no longer consider the text. If a user believes a reading material fits the context but was not recommended, they have the option to manually click ``Add Text'' to request the system to analyze it based on the context.
To optimize generation speed, the system generates a limited number of contexts and analyzes a limited number of reading materials at a time (set as 8 during experiments). Users can click ``More Contexts'' and ``More Contexts'' to generate additional contexts and analyze more texts for further exploration.

\begin{figure*}[]
  \centering
  \includegraphics[width=\linewidth]{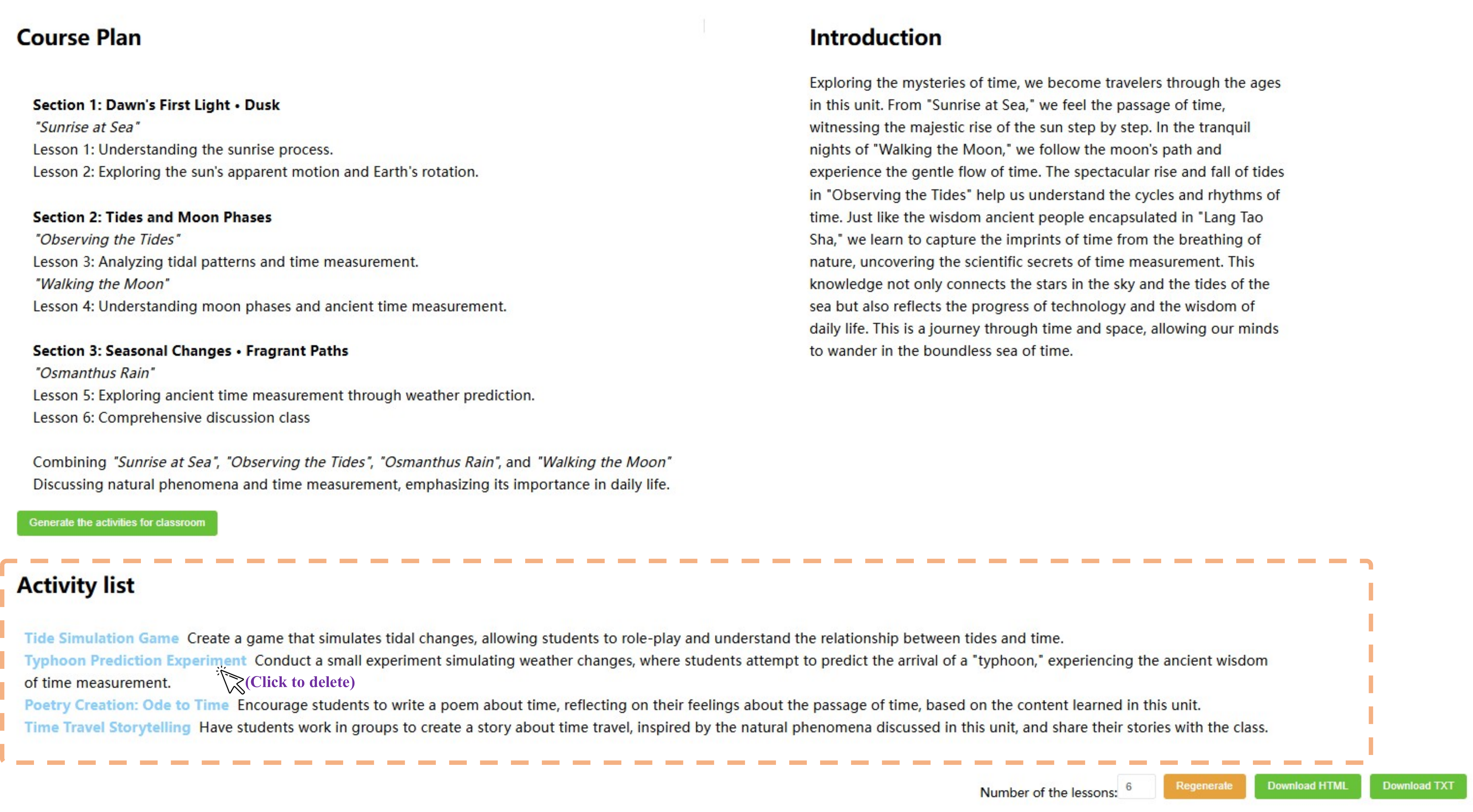}
  \caption{The user interface of \name{}'s Outcome Generation Panel, translated from Chinese either by Google Translator or manually. The content in this figure serves as an example of outcomes.}
  \Description{The user interface of \name{}'s Outcome Generation Panel, translated from Chinese either by Google Translator or manually. The content in this figure serves as an example of outcomes.}
\label{fig:system_interface2}
\end{figure*}

When clicking a context within Collection View  \pzh{(\autoref{fig:system_interface})}, users are directed to the Outcome Generation Panel \pzh{(\autoref{fig:system_interface2})}.
In this panel, users can examine the details of their chosen context as well as the analyses of the associated reading materials. 
Subsequently, users are prompted to enter the ``Expected number of lessons'' (\autoref{fig:system_interface} G) and select the ``Generate Introduction and Course Plan'' (\autoref{fig:system_interface2}) to create a comprehensive course plan that includes explanations for each segment and an introduction aligned with the course plan. 
In this interface, all content can be edited by double-clicking to adjust the final outcomes. Clicking ``Generate the Activities for Classroom'' recommends teaching activities related to the literature and the interdisciplinary subjects. Users can click on the titles to delete unnecessary activities.
At this point, a complete outcome based on a specific context has been generated.
Clicking the download button at the bottom right of the panel allows users to download the content in txt and HTML formats for sharing or further editing.

\subsection{Implementation of the System}
\begin{figure*}[]
  \centering
  \includegraphics[width=0.90\linewidth]{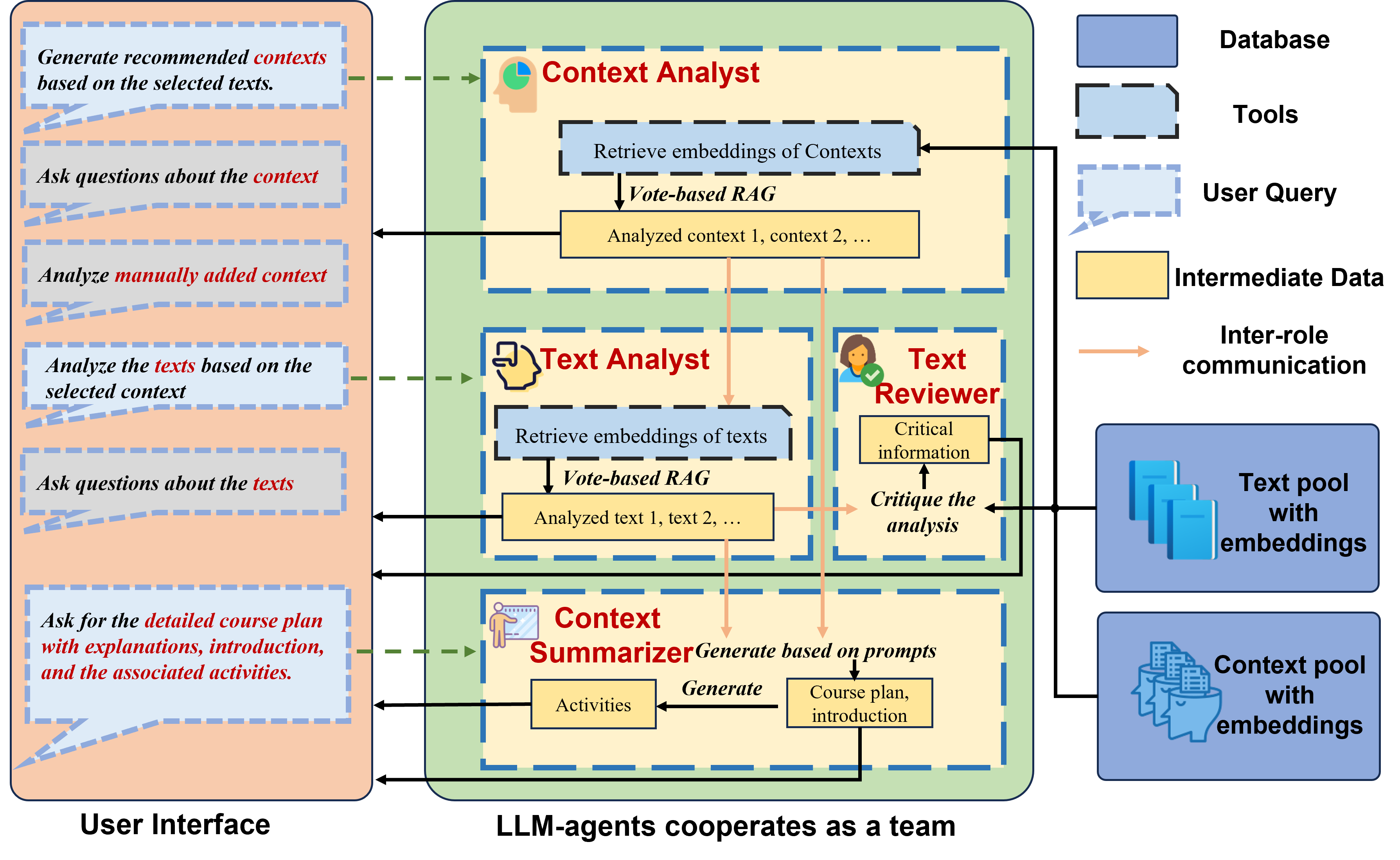}
  \caption{\fhx{The architecture of \name{}. The illustration of the main workflow of \name{}. The responses to user's queries with gray background are not shown in this figure.}}
  \Description{The architecture of \name{}. The illustration of the main workflow of \name{}. The responses to user's queries with gray background are not shown in this figure.}
\label{fig:implement}
\end{figure*}

\fanhx{
Inspired by~\cite{hou2024c2ideas, wu2024transagents}, \name{} aligns its behavior with teachers' habitual practices in interdisciplinary ideation (DP1).
We designed an LLM-based multi-agent system to simulate human collaboration.
Following the multi-agent collaborative framework by~\citet{hong2024metagpt}, we decompose the task into multiple steps, allowing agents to use tools when necessary and interact with the user for feedback.
Also, the prompt design follows the Co-Star framework~\cite{napoli2024leveraging} to optimize outputs.
}
The architecture of our workflow is illustrated in~\autoref{fig:implement}, which will be discussed in~\autoref{sbsbsc:architecture}.
The effectiveness of the content generated through this architecture relies on a comprehensive Context Pool (DP4) and carefully designed prompts (DP2, DP3), which will be introduced in~\autoref{sbsbsc:contextpool} and~\autoref{sbsbsc:Iterativeprompt}, respectively.

\subsubsection{Architecture of \name{}}
\label{sbsbsc:architecture}
\penguin{In line with the teachers' habitual practices and DPs,} 
we simulate four roles \fanhx{(DP1)} that participate in the ideation of interdisciplinary contexts: Context Analyst \penguin{(DP2)}, Text Analyst \penguin{(DP2)}, Text Reviewer \penguin{(DP2, DP4)}, and Context Summarizer \penguin{(DP3)}.
To facilitate their collaboration, similar to MetaGPT~\cite{hong2024metagpt}, we enable these agents to remember, think (\ie determine subsequent actions based on observations), act (\ie utilize tools and communicate with one another), and, crucially, observe user behavior through the interface to respond appropriately.
Each role is equipped with the GLM4 model as its ``brain.''
Before the application is implemented, all original contexts and reading materials are embedded using the Zhipu embedding-3 model\footnote{\url{https://bigmodel.cn/dev/api/vector/embedding-3}}. The texts and embedding arrays are stored in the database.
For contexts imported by users, \name{} will complete this embedding process in real-time.

\textbf{Context Analyst} acts as a ``teacher with expertise across various disciplines.'' When a user clicks ``Generate Recommended Contexts,'' it retrieves background information of the original contexts that are most relevant to the reading materials from the database
. For each context, Context Analyst conducts an analysis based on all associated reading materials and generates a detailed description of the context. The Context Analyst's memory continuously retains the texts and analyzes contexts, similar to the role of a teacher. When a user initiates a request in ``Find'' mode, Context Analyst is tasked with providing detailed information related to the context.

\textbf{Text Analyst} acts as an ``active-minded literature teacher.'' When a user requests an analysis of texts based on the selected context, Context Analyst provides Text Analyst with the details of the selected context. Initially, Text Analyst
retrieves the most relevant texts. Subsequently, Text Analyst conducts a thorough analysis of each text individually, producing an in-depth single-text analysis. Furthermore, Text Analyst responds to users' inquiries related to the text.

\textbf{Text Reviewer} acts as a ``conservative and experienced literature teacher.'' When Text Analyst generates an analysis, it seeks Text Reviewer's opinion, which entails Text Reviewer assessing the relevance and accuracy of the analysis based on the reading material and the selected context, providing evaluative information to the user. This additional review process provides users with critical information and corrects potential inaccuracies in the RAG.

\textbf{Context Summarizer} acts as a ``literature teacher skilled in summarization,'' responsible for integrating the context with synthesizing contextual information with selected reading materials from the collection. It delivers structured outputs step-by-step in the Outcome Generation Panel (\ie a detailed lesson plan with explanations for each segment, the introduction based on the lesson plan, and the associated activities).

In summary, various user actions within the interface trigger the actions of different roles. Upon reflection, these roles employ a range of tools and capabilities, which depend on carefully designed prompts, to accomplish their respective tasks.

\subsubsection{Context Pool}
\label{sbsbsc:contextpool}
To enhance the generation of reasonable and evidence-based contexts, we compiled a database of original contexts accompanied by background information across various subjects (including general contexts). 

The Informal Education Database contains 113 original contexts derived from a famous Chinese journal Yao Wen Jiao Zi's ``Annual Popular Words (2008-2023) \footnote{\url{http://yaowenjiaozi.cn/}}.''
These materials claim to be based on data from the National Language Resources Corpus, with additional content selected manually. 
One author read the 160 original contexts and included those comprehensible to elementary school students in the database. 
The context database for the subjects of art, science, mathematics, and music is extracted from the unit titles and descriptions of widely used textbooks for grades 3-6, including 144 original texts.
These contexts were selected to fulfill teachers' requirements for accuracy and diversity, providing a positive experience during the user study. Notably, teachers can easily add contexts in a batch; they only need to provide the subject, title, and background information, similar to the operations performed on the interface.

\subsubsection{Iterative Prompt Engineering}
\label{sbsbsc:Iterativeprompt}
\name{} aims to provide users with high-quality, structured outcomes based on specific evaluation criteria. We have incorporated valuable insights from experts gathered during the foundational study into the prompt design. The prompts are structured in accordance with the co-star~\cite{napoli2024leveraging} framework, which encompasses context, objective, style, tone, audience, and response in a single prompt. The style, tone, and audience specifications for each role are shown in the supplementary materials.
The design details of the context, objective, and response are shown below.

The \textbf{context part} of the prompt not only explains the task (\eg find multiple points related to the context across various texts and generate a highly specific analysis based on the context) but also provides all the requisite knowledge for the task. Although the GLM-4 supports a 128k context length, overly long prompts may lead the LLM to neglect certain information. Therefore, we precisely designed the minimum information necessary for each role's actions. For instance, Text Analyst is provided solely with the content of the text to be analyzed and the relevant context for each request, and this procedure is repeated for multiple texts to ensure the most precise analysis for each individual text. Meanwhile, Context Summarizer concentrates on generating relevant activities in accordance with the course plan, incorporating only the course plan and introduction into the prompt, without the content of the reading materials.

The \textbf{objectives part} in all prompts are consistently aligned with the six metrics established by the team (DP3) for evaluating the quality of contexts and outcomes. These clearly defined objective statements ensure that the outputs are in close accordance with the expectations of teachers. The response, which specifies the required output format and content, is embedded with templates developed by experts during the foundational study. Following the evaluation of the prototype, we provided feedback to experts E1-E6 and revised the templates to enhance their content richness. The template for Context Summarizer for generating a course plan is as follows:

\begin{quote}
Segment 1: Initial Encounter with the Scene · Uncontrollable Emotions
\begin{itemize}
    \item ``Prairie'' [the title of the text] (The natural scenery and ethnic friendship in the prairie)
    \begin{itemize}
        \item Lesson 1: Appreciate the beautiful sentences in the text and feel the beauty of the prairie.
        \item Lesson 2: Feel the enthusiasm of the prairie people and the deep friendship between Mongolian and Han people.
    \end{itemize}
\end{itemize}

Segment 2: Encounter with the Scene Again · Touching the Heartstrings
\begin{itemize}
    \item ``Lilac Knot'' [the title of the text] (The feelings in the lilac knot)
    \begin{itemize}
        \item Lesson 3: Understand the author's way of expressing associations triggered by objects and explain your understanding.
        \item Lesson 4: Understand the symbolic meaning of the lilac knot and appreciate the emotions embedded by the author.
    \end{itemize}
    \item ``Lodging by the River'' [the title of the text] + ``Visiting an Old Friend'' [the title of the text] (The feelings in the mountains and rivers under the moonlight)
    \begin{itemize}
        \item Lesson 5: Use ``Lodging by the River'' as an example to teach the method of learning ancient poetry and explore the imagery of the ``moon.''
        \item Lesson 6: Imagine the scenes in ``Lodging by the River'' and ``Visiting an Old Friend'' and understand the unique poetic feelings evoked by the ``mountains and rivers.''
    \end{itemize}
\end{itemize}

Segment 3: Another Encounter with the Scene · Endless Thoughts
\begin{itemize}
    \item ``Song of Flowers'' [the title of the text] (Unity of man and nature, understanding philosophy)
    \begin{itemize}
        \item Lesson 7: Feel the attitude towards life and inner ideals after transforming into a ``flower,'' and appreciate the author's positive attitude towards life.
    \end{itemize}
\end{itemize}
\end{quote}

The template includes all the elements expected by teachers during the foundational study. The LLM can utilize the template to verify that the response includes all necessary content.

\section{Experiment I: Within-subject study}\label{sec:experiment_1}
\pzh{We evaluate \name{} via two experiments. Experiment I is a within-subjects study that aims at assessing the effectiveness of \name{} in supporting novices of literature teaching to come up with interdisciplinary contexts. 
Experiment II is a qualitative study that examines \name{}'s values in assisting expert Chinese language teachers in ideating interdisciplinary contexts for literature teaching. 
}
In this section, we present the design and results of Experiment I.
\fanhx{
The research questions are:
\begin{itemize}
    \item RQ1. How would \name{} affect the interdisciplinary exploration outcomes?
    \item RQ2. How would \name{} affect the process of interdisciplinary exploration?
    \item RQ3. How would users perceive the usability and creative support of \name{}?
\end{itemize}
}

\subsection{Participants}
We recruited 16 students (P1-P16, nine females, seven males; age: \textit{Mean}=23.44, \textit{SD}=2.24) via a post in a group chat and via word-of-mouth from two universities. 
\penguin{This sample size follows the practices of previous work that evaluates systems for supporting ideation and education, such as CreativeConnect (N = 16) \cite{choi2024creativeconnect}, AngleKindling (N = 12) \cite{petridis2023anglekindling}, DiaryHelper (N = 12) \cite{li2024diaryhelper}, and LessonPlanner (N = 12) \cite{fan2024lessonplanner}. 
We also calculated the required sample size with G* Power to conduct Wilcoxon Signed-Rank tests that compare measures in \name{} and baseline conditions. 
With Tails = Two (can not tell in advance which of the two measures is larger), Parent distribution = Normal (default), Effect size = 0.80 (calculated as Cohen's d~\cite{cohen2013statistical}, large effect), $\alpha \text{ err prob} = 0.05$ (default), and power $(1 - \beta \text{ err prob}) = 0.8$ (an acceptable threshold), the output recommended smallest sample size is 15. 
}
\pzh{Five participants, including two third-year undergraduate and three graduate students, major in}
\penguin{the fields related to computer science (CS)}. 
\penguin{They could provide feedback from those who know about the techniques used in \name{}, and they were capable of our ideation tasks as they all learned Chinese literature in elementary schools.} 
The other eleven participants are all graduate students majoring in education-related fields. \penguin{All of them are teachers in training, while P2, P4, and P9 have teaching experience. The others are pre-service teachers.}. 
All participants expressed their interest in utilizing web resources and AI-assisted tools to design interdisciplinary contexts (\textit{Mean} = 4.62, \textit{SD} = 0.48; 1 - no interest at all, 5 - very interest in). All participants reported having experience with large language models (LLMs, \eg ChatGPT) (\textit{Mean} = 4.0, \textit{SD} = 0.94; 1- no experience, 5 - use daily).

\begin{table*}[h]
\caption{Participants involved in the within-subjects study}
\Description{Participants involved in the within-subjects study}
\begin{tabular}{@{}cccccc@{}}
\toprule
\textbf{ID} & \textbf{Gender} & \textbf{Year} & \textbf{Age} & \textbf{Major}                                         & \textbf{Freq. of AI Usage} \\ \midrule
P1  & F & Graduate      & 23 & Curriculum and Instruction – English Language Education & Daily        \\
P2  & M & Graduate      & 23 & Curriculum and Instruction – English Language Education & Weekly       \\
P3  & F & Graduate      & 23 & Curriculum and Instruction – English Language Education & Weekly       \\
P4  & F & Graduate      & 25 & Curriculum and Instruction – English Language Education & Weekly       \\
P5  & F & Graduate      & 24 & Curriculum and Instruction – English Language Education & Daily        \\
P6  & F & Graduate      & 23 & Chinese Literature                                      & Have Tried   \\
P7  & F & Graduate      & 22 & Early Childhood Education and Child Development         & Weekly       \\
P8  & F & Graduate      & 22 & Early Childhood Education and Child Development         & Infrequently \\
P9  & F & Graduate      & 28 & Educational Psychology                                  & Daily        \\
P10         & F               & Graduate      & 23           & TESOL(Teaching English to Speakers of Other Languages) & Have Tried                 \\
P11 & M & Graduate      & 29 & Computer Science and Technology                         & Weekly       \\
P12 & M & Graduate      & 22 & Computer Science and Technology                         & Daily        \\
P13 & M & Graduate      & 23 & Electronics and Information Engineering                 & Weekly       \\
P14 & M & Undergraduate & 20 & Artificial Intelligence                                 & Weekly       \\
P15 & M & Undergraduate & 21 & Artificial Intelligence                                 & Weekly       \\
P16 & M & Graduate      & 24 & Education                                               & Daily        \\ \bottomrule
\end{tabular}
\end{table*}

\subsection{Experimental Design}
Experiment I is a within-subject design. Each participant completed one interdisciplinary exploration task with \name{} and the other with a baseline setup.

\subsubsection{Baseline Condition and \name{} Condition}
The baseline system is the LLM web application, which employs the Zhipu LLM model, the same model implemented in \name{}. 
Participants could upload and read document files, thereby enabling the web app to access selected reading materials. 
Also, the participants had the option to utilize the Google search engine and Microsoft Office Word. 
In the \name{} condition, participants were allowed to use Google, Microsoft Office Word, and \name{}, but were not allowed to use the LLM web app.
In both conditions, participants could freely choose whether, when, and how to utilize the provided tools.

\subsubsection{Tasks-systems Assignment}
\pzh{We invited the experienced Chinese language teacher E1 (\autoref{tab:teachers}), who participated in our design process, to help us prepare the task materials and grade the task outcome}. 
Specifically, E1 selected 16 reading materials from six textbooks, covering 15 different units, and divided them into two groups - Text Set 1 and Text Set 2.
Each participant completed tasks with both groups of texts. The task assignments were as follows:
\begin{itemize}
    \item Baseline (Text Set 1) + \name{} (Text Set 2)
    \item \name{} (Text Set 2) + Baseline (Text Set 1)
    \item \name{} (Text Set 1) + Baseline (Text Set 2)
    \item Baseline (Text Set 2) + \name{} (Text Set 1)
\end{itemize}

\subsubsection{Procedure}
Each participant was assigned two lesson-planning tasks.  
The task prompt was: 
\begin{quote}
    You are a sixth-grade Chinese language teacher, aiming to implement an interdisciplinary approach over a period of 3-5 days, establishing logical and thematic connections between in-class and extracurricular reading materials. You are facing challenges in organizing these reading materials. Your goal is to analyze the internal relationships among the texts, identify an `interdisciplinary context', integrate the selected texts within this context, develop a course plan in a document explaining the relationship between the reading materials and the selected context, and propose potential literature or interdisciplinary instructional activities. 
    You may refer to the provided documents \pzh{of example lesson plans based on interdisciplinary contexts}. 
    \pzh{You need to choose at least three from eight reading materials and integrate them with the context.}
\end{quote}

Participants received the original reading materials one day before the task and were instructed to spend 10 minutes reading them.
On the day of the task, the four participants assigned to the same task-system group came to our lab. 
We introduced the background knowledge about literature and interdisciplinary instruction, and the components necessary for creating an effective context (\ref{sec:principles} DP2).
To encourage their best performance, we informed the participants that their outputs would be evaluated by an expert based on the appropriateness of Context, alignment with the educational objectives, and depth of integration. The top three participants would be awarded a supplementary reward of 100 RMB.

For each task, we first introduced the task and demonstrated the assigned system (\name{} or baseline).
Subsequently, we assisted participants in setting up their \pzh{task} environment on the provided computers or their own devices, including opening the baseline system or \name{}, Google, and Microsoft Office Word. 
Each participant independently completed the lesson planning task. 
We allocated 30 minutes for each task and informed them that they could complete the task ahead of time if they wished. 
Following the completion of each task, participants were required to fill out a questionnaire. A 10-minute break was provided between the two tasks.  
After completing both tasks, we conducted a final semi-structured interview. Each participant spent approximately 90 minutes in the experiment and received 100 RMB for compensation.

\subsection{Measurements}

\penguin{
\textbf{RQ1. Ideation outcomes.}} 
To assess the quality of the outcomes, we invited E1 to rate all outcomes in a randomized order.
\penguin{E1 did not know which condition each output lesson plan came from and reported that she spent four to six minutes evaluating each outcome, and} 
the assessment was based on three criteria 
\penguin{(derived in Finding 6 in \autoref{sec:principles})}
: Content Alignment, Alignment with Educational Objectives, and Depth of Integration, using a 7-point Likert scale \penguin{(1 - not satisfied at all, 7 - fully satisfied)} to indicate the extent to which each criterion was satisfied. Additionally, \penguin{to make sense of the ratings, we required E1 to} provide comments for each lesson plan, highlighting good and bad aspects.

\penguin{
\textbf{RQ2. Ideation process.}} Based on NASA-TLX, we formulated six questions to measure workload during the lesson planning process, including mental demand, physical demand, temporal demand, performance, effort, and frustration - \penguin{the higher a TLX score suggests the higher the perceived workload.}

\penguin{
\textbf{RQ3. Perception of \name{}.}} We adapted ten questions from the System Usability Scale (SUS)~\cite{brooke2013sus} to study effectiveness, efficiency, and satisfaction. 
To understand how users perceive \name{}'s generated contexts and analysis, we adapted six questions from the Creativity Support Index (CSI)~\cite{cherry2014quantifying} to evaluate Exploration, Expressiveness, and Immersion. \yh{The Creativity Support Index (CSI) includes three key dimensions: Exploratory, which measures how well the system helps users explore diverse ideas; Expressiveness, which reflects the system’s ability to support clear expression of creative ideas; and Immersion, which indicates how well the system allows users to stay focused on creative tasks.}

\pzh{The measured items for task workload, SUS, and CSI are rated using a standard 7-point Likert Scale (1 - strongly disagree, 7 - strongly agree).}

\subsection{Results and Analyses}
For quantitative data, we used the Wilcoxon Signed-Rank test.
\penguin{
We used G* Power software to compute sensitivity, given Parent distribution = Normal, $\alpha = 0.05$, Power = 0.8, and Total Sample size = 16, which outputs a required effect size of 0.77. 
}
For qualitative data,
\fanhx{
we conducted thematic analysis on the semi-structured interview under the same settings in the design process (\autoref{sec:formative_findings}) and presented comments from the outcomes as supporting evidence.
}

\subsubsection{RQ1: Ideation Outcomes}
\begin{figure}[]
  \centering
  \includegraphics[width=1\linewidth]{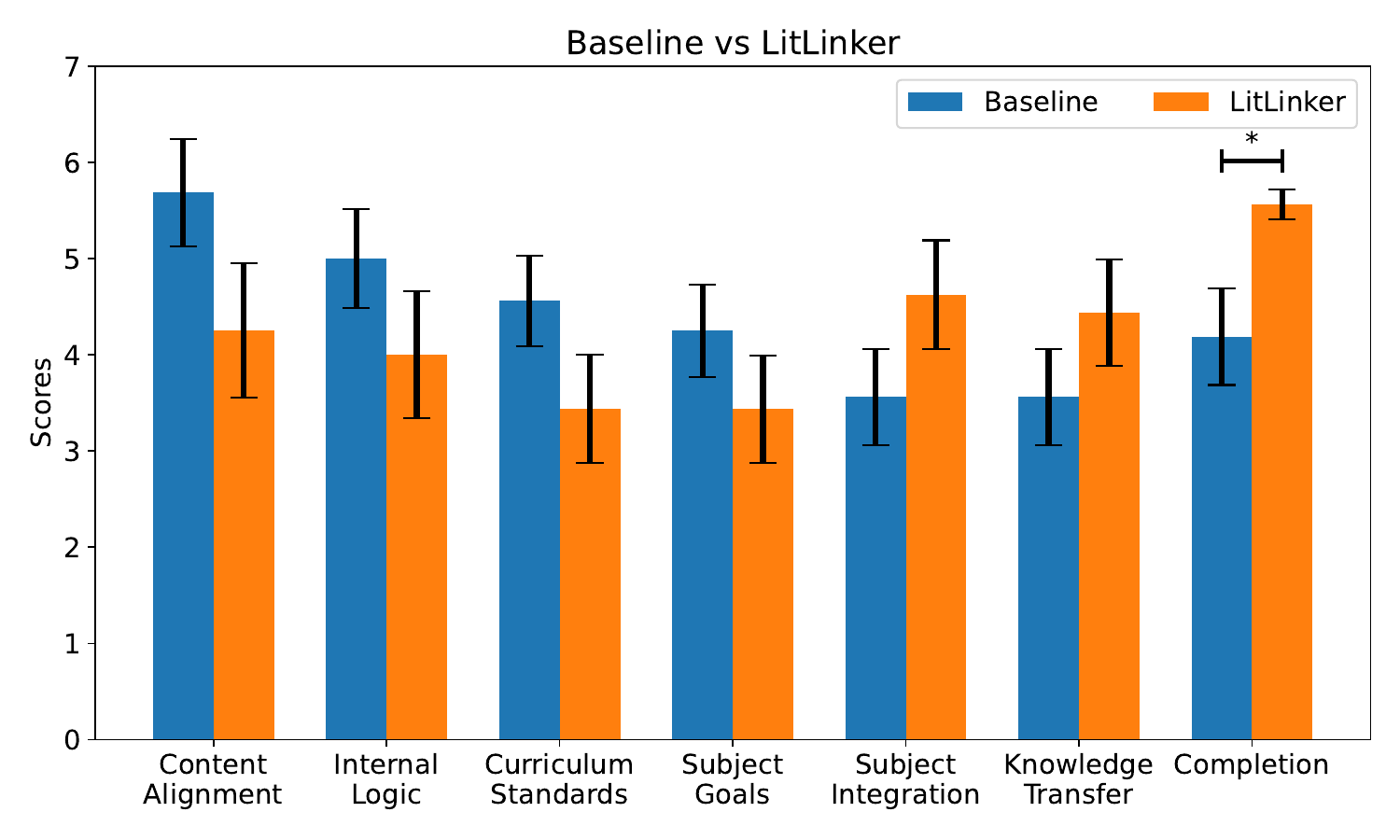}
  \caption{RQ1 results regarding the outcomes evaluated by E1 in seven different aspects. ***: p<0.001, **: p<0.01, *: p<0.05, +: p<0.1}
  \Description{RQ1 results regarding the outcomes evaluated by E1 in seven different aspects. ***: p<0.001, **: p<0.01, *: p<0.05, +: p<0.1}
  \label{fig:rq1_results}
\end{figure}

\autoref{fig:rq1_results} shows the quality of outcome lesson plans from the within-subjects study.
The results indicate high variance in outcome quality for both the baseline and \name{} conditions, suggesting significant differences in user performance. 
\name{} showed improvements in Completion (\name{}: $M = 5.56, Mdn=6, SD = 0.61$; baseline: $M = 4.19, Mdn=4.5, SD = 2.01$; $p = 0.02$\fhx{, Wilcoxon effect size ($r$) = 0.93}), providing more guidance in organizing lesson components. In Subject Integration (\name{}: $M = 4.63, Mdn=6, SD = 2.26$; baseline: $M = 3.56, Mdn=3, SD = 2.00$; $ p = 0.34$\fhx{, $r = 0.50$}), \name{} assisted in combining literature knowledge with disciplinary contexts like science and arts. For Knowledge Transfer (\name{}: $M = 4.44, Mdn=5.5, SD = 2.21$; baseline: $M = 3.56, Mdn=3, SD = 2.00$; $p = 0.38$\fhx{, $r = 0.42$}), outcomes created with \name{} were more likely to promote students' application of disciplinary knowledge in their classroom environment.
\pzh{These results suggest that \name{} performed well in supporting users to connect the interdisciplinary contexts to the reading materials.}

However, regarding Content Alignment (\name{}: $M = 4.25, Mdn=5, SD = 2.80$; baseline: $M = 5.69, Mdn=7, SD = 2.23$; \pzh{$p = 0.22$}\fhx{, $r = 0.57$}), Internal Logic (\name{}: $M = 4.00, Mdn=3.5, SD = 2.65$; baseline: $M = 5.00, Mdn=5.5, SD = 2.06$; \pzh{$p = 0.29$}\fhx{, $r = 0.42$}), Subject Objectives (\name{}: $M = 3.44, Mdn=3, SD = 2.24$; baseline: $M = 4.25, Mdn=5, SD = 1.89$; \pzh{$p = 0.32$}\fhx{, $r = 0.39$}), and Curriculum Standards (\name{}: $M = 3.44, Mdn=3, SD = 2.65$; baseline: $M = 4.56, Mdn=6, SD = 1.93$; \pzh{$p = 0.16$}\fhx{, $r = 0.53$}), \pzh{the outcome lesson plans with the baseline system were rated significantly higher than those with \name{}.}
This indicates that the outputs from \name{} were less appropriate in terms of \pzh{aligning the contexts} with educational objectives compared to the baseline. 
\pzh{We identify two possible reasons for these results based on the experimental setup and E1's comments on the outcome lesson plans.}
First, few participants had teaching experience, \pzh{and none of them had taught Chinese literature before}.
\pzh{In other words, participants were} unfamiliar with the specific educational objectives associated with literature. 
During the introduction of the system and the task, we overemphasized the concept of ``interdisciplinary'', which led participants to focus more on other subjects rather than the educational objectives of the literature itself. In contrast, the experienced teacher E1 who evaluated the outcomes is highly sensitive to curriculum standards and literature objectives. E1 commented on an outcome that received a Curriculum Standards score of 1 (P08 - \name{}), \textit{``The forced interdisciplinary integration is counterproductive, as it shifts the focus from literature to art or science, losing sight of the primary subject.''}
Second, the system occasionally provided excessive or inaccurate interpretations of the texts, leading to lower appropriateness of context scores. \textit{``I feel that the system's trying to fit the analysis into a specific context, but it does not always feel very relevant.''} (P7). 
Most participants were not familiar with the reading materials, making it difficult for them to identify these issues. In contrast, E1 could easily identify such deviation. E1 commented on an outcome that received an appropriateness of context score of 1 (P11 - \name{}), \textit{``The analysis does not match my understanding of the text.''}

In summary, while \name{} improved Completion, Subject Integration, and Knowledge Transfer \pzh{of the outcome lesson plans}, it resulted in lower ratings concerning the appropriateness of context and alignment with educational objectives. 
This might be due to the experimental setup and the system's misinterpretation of the texts. We will focus on these issues in expert interviews to evaluate the quality of the outcomes.

\subsubsection{RQ2: Ideation Process}
\begin{figure}[]
  \centering
  \includegraphics[width=1\linewidth]{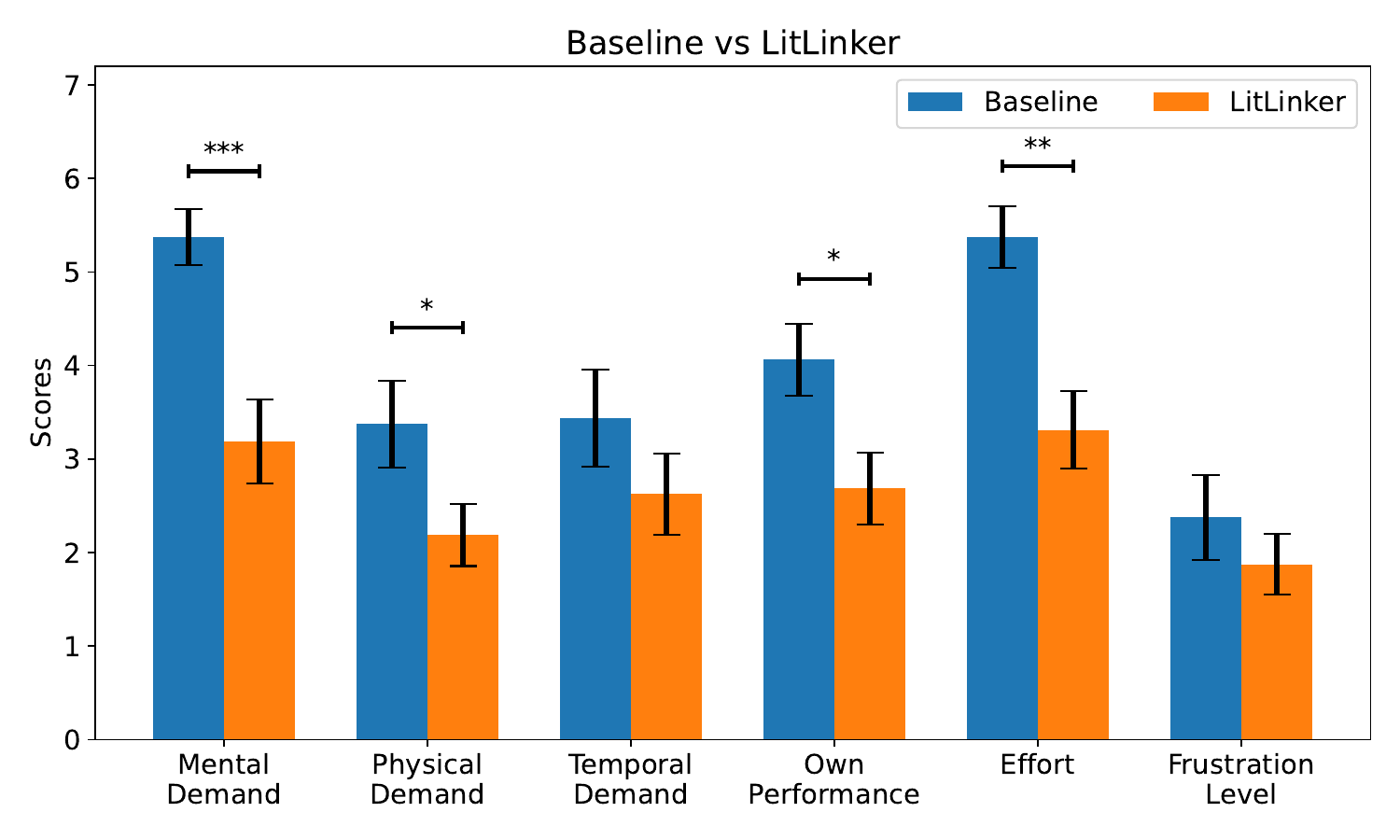}
  \caption{
  \penguin{RQ2 results regarding evaluating how \name{} affects the workload during the task. For each metric, a higher NASA-TLX score suggests a higher perceived workload. ***: p<0.001, **: p<0.01, *: p<0.05, +: p<0.1}}
  \Description{RQ2 results regarding evaluating how \name{} affects the workload during the task. For each metric, a higher NASA-TLX score suggests a higher perceived workload. ***: p<0.001, **: p<0.01, *: p<0.05, +: p<0.1}
  \label{fig:rq3_results}
\end{figure}

\autoref{fig:rq3_results} shows the participants' ratings on six NASA-TLX dimensions between the baseline system and \name{} in the user study. 
Users rated mental demand and physical demand significantly lower with \name{} (\name{}: $M = 3.19, Mdn=3, SD = 1.74$; baseline: $M = 5.38, Mdn=5, SD = 1.17$, $p=0.0009$\fhx{, $r = 1.48$}; \name{}: $M = 2.19, Mdn=2, SD = 1.29$; baseline: $M = 3.38, Mdn=2.5, SD = 1.80$, $p=0.01$\fhx{, $r = 0.76$}). Users felt significantly more satisfied with their performance using \name{} (\name{}: $M = 2.69, Mdn=2, SD = 1.49$; baseline: $M = 4.06, Mdn=4.5, SD = 1.48$; $p = 0.03$\fhx{, $r = 0.93$}) and reported significantly lower Effort (\name{}: $M = 3.31, Mdn=3.5, SD = 1.61$; baseline: $M = 5.38, Mdn=5.5, SD = 1.27$; $p = 0.003$\fhx{, $r = 1.42$}). Frustration Level scores were also significantly lower with \name{} (\name{}: $M = 1.88, Mdn=1.5, SD = 1.27$; baseline: $M = 2.38, Mdn=1, SD = 1.76$; $p = 0.23$\fhx{, $r = 0.33$}), suggesting that users felt more confident during the exploration process. Additionally, users reported lower average Temporal demand with \name{} (\name{}: $M = 2.63, Mdn=2, SD = 1.69$; baseline: $M = 3.44, Mdn=3, SD = 2.00$; $p = 0.27$\fhx{, $r = 0.44$}). 
\pzh{In summary, the participants indicate that \name{} alleviates workload compared to the baseline system across all dimensions.}

\subsubsection{RQ3: Perception of \name{}}
\begin{figure}[]
  \centering
  \includegraphics[width=1\linewidth]{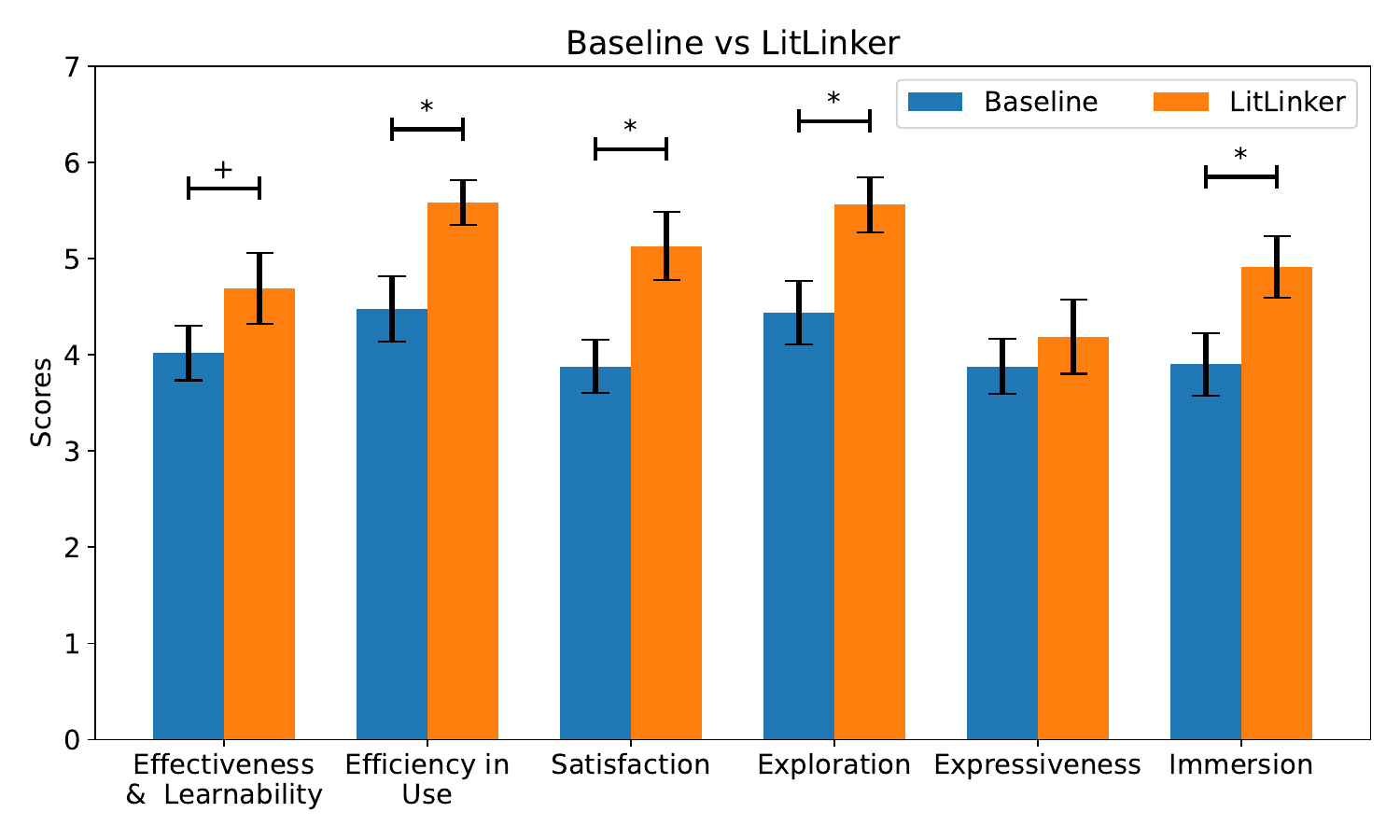}
  \caption{
  \fhx{RQ3 results regarding users' perception of \name{}, based on 3 aspects from SUS, and 3 aspects from CSI. ***: p<0.001, **: p<0.01, *: p<0.05, +: p<0.1}}
  \Description{RQ3 results regarding users' perception of \name{}, based on 3 aspects from SUS, and 3 aspects from CSI. ***: p<0.001, **: p<0.01, *: p<0.05, +: p<0.1}
  \label{fig:rq2_results}
\end{figure}

\fhx{\autoref{fig:rq2_results}} presents the statistical results comparing the baseline system and \name{} based on effectiveness, efficiency, and satisfaction scores. 
Participants rated \name{} higher in all three aspects. Satisfaction scores were significantly higher for \name{} (\name{}: $M = 5.13, Mdn=5.5, SD = 1.42$; baseline: $M = 3.88, Mdn=3.83, SD = 1.10$; $ p = 0.03$\fhx{, $r = 0.99$}), indicating greater user satisfaction with \name{}. Efficiency scores also showed a significant increase with \name{} (\name{}: $M = 5.58, Mdn=5.63, SD = 0.93$; baseline: $M = 4.48, Mdn=4.75, SD = 1.36$; $p = 0.02$\fhx{, $r = 0.94$}), suggesting that participants found \name{} more helpful and relevant to their tasks. 
For effectiveness, \name{} received a higher mean score (\name{}: $M = 4.69, Mdn=4.83, SD = 1.48$; baseline: $M = 4.02, Mdn=4, SD = 1.14$; $ p = 0.06$\fhx{, $r = 0.50$}), implying that participants felt more effective using \name{}.

For the three dimensions of the Creativity Support Index (Exploratory, Expressiveness, and Immersion), our system achieved higher average scores in all aspects. Users rated \name{} significantly higher on Exploratory dimension (\name{}: $M = 5.56, Mdn= 5.75, SD = 1.14$; baseline: $M = 4.44, Mdn=4.5, SD = 1.31$; $p = 0.01$\fhx{, $r = 0.91$}), suggesting better support for exploratory creativity. On the Immersion dimension, users also gave \name{} significantly higher ratings (\name{}: $M = 4.91, Mdn=5, SD = 1.28$; baseline: $M = 3.90, Mdn=4.5, SD = 1.30$;$ p = 0.02$\fhx{, $r = 0.75$}), indicating an enhanced ability to focus during creative tasks. For the Expressiveness dimension, users rated \name{} higher (\name{}: $M = 4.19, Mdn=4.5, SD = 1.55$; baseline: $M = 3.88, Mdn=3.75, SD = 1.14$; $p = 0.36$\fhx{, $r = 0.23$}), indicating a moderate improvement in expressiveness.
\peng{Despite the overall strengths of \name{} in creative support, six participants in the semi-structured interviews pointed out ways to further improve its creative support.}
First, in terms of exploratory dimensions, the system's interpretation of contexts and the quality of the AI-generated responses did not align with the expectations of some users, thereby diminishing their motivation to engage in further exploration. \textit{``(For a science context,) I expected it to provide more information related to students' everyday lives, but it mostly analyzed from abstract directions like aesthetics, which didn't match my expectations.''} (P6). \textit{``Its answers didn't address the questions I wanted to explore; sometimes it felt like it was using fancy language to evade the core issues.''} (P13). 
Regarding Expressiveness, \pzh{four} participants noted that \name{} tended to ``persuade'' them to accept the contexts established by the system, rather than supporting their own critical thinking and expression. 
\textit{``It felt like scrolling through TikTok, reading the AI-generated decent outcomes without much thought.''} (\pzh{P12}).
\penguin{
We observed that the eleven participants with education-related backgrounds ($M = 3.77, SD = 1.59$) tended to give lower scores on Expressiveness than the five students ($M = 5.10, SD =1.34$) with CS-related backgrounds. 
P7, a student from an educational major, stated that 
\textit{
``I find the system too acquiescent; it often responds with `You are right' or `I agree with you'. However, sometimes my thoughts were not necessarily correct, and I would prefer it to provide a more critical perspective.''
}
This suggests the system should further customize the support to users with education majors to construct and refine their initial ideas. 
}



\section{Experiment II: Expert Interviews} \label{sec:experiment_2}
In Experiment I, we evaluated novice teachers' perceptions of our system, including the ideation process and their overall experience using \name{}.
Additionally, an experienced teacher, E1, evaluated the outcomes of the novice teachers. In Experiment II, we shifted our focus to the perspectives of experts \pzh{with varying levels of literature teaching experiences} to gain more feedback. 
We conducted expert interviews using a think-aloud protocol and a semi-structured interview with nine teachers to collect insights related to our research questions. 

\subsection{Participants}
The study involved nine \pzh{Chinese language} teachers (5 female, 4 male, E1-5, E8-11 in \autoref{tab:teachers}) in a local elementary school, 
including two novice teachers with less than five years of teaching experience and six expert teachers with at least five years of teaching experience~\cite{booth2021mid}.
Detailed information about the participants is presented in Table 3.

\subsection{Method}
\penguin{
We conducted interviews offline with E8-E11 and I1-I6 in lab sessions and with E1-E5 online after they freely used it for three days. This setup could help us to gain diverse insights, \eg learnability of \name{} in lab and field environments. 
I1-I6 participated in the evaluation of \name{}'s prototype, and their feedback on the unchanged features in \name{} was also presented in this section. 
}

For \textbf{E8 - E11}, who were not involved in the iterative design process, we conducted offline interviews. The process began with a 5-minute introduction to the research background (\ie the background of interdisciplinary literature instruction). 
This was followed by a 10-minute tutorial on how to use \name{}. 
Participants were then allocated 30 minutes to complete exploration tasks while engaging in a think-aloud protocol. Subsequently, a 15-minute semi-structured interview was conducted.

For \textbf{I1 - I6}, who participated in the evaluation of the prototype process, the procedure was consistent with the same methodology above, except that during the semi-structured interview, we asked for more suggestions regarding the prototype.

For \textbf{E1 - E5}, who engaged in the iterative design process, we conducted online interviews. The \name{} was made available for a duration of 3 days, and a tutorial video was provided. Participants were asked to freely use \name{} during this period to complete exploration tasks. Finally, we conducted a 15-minute semi-structured interview with each expert.

\fanhx{
Surrounding our RQs, the interview questions (\autoref{sec:appendix}) are about ideation outcomes, ideation process and perception of \name{}, while the participants were assigned the following exploration tasks:
\begin{itemize}
    \item Task 1: Freely choose 4-15 reading materials, explore two contexts of interest, and add them to the collection.
    \item Task 2: Generate an introduction, course plan, and activities.
\end{itemize}
}




\subsection{Findings}
\subsubsection{RQ1: Ideation Outcomes}
Overall, seven of nine teachers confirmed that the outcomes generated by \name{} effectively support literature instruction within the classroom environment. E10 emphasized the comprehensiveness of the text analysis,
\penguin{where she could find the content she wanted within a document more strategically, instead of sifting through countless reference books:} 
\penguin{\textit{``I can abandon varied reference books, (because) the system conducts a comprehensive analysis of the specific content of the texts. It includes themes, content, and key points.''}}
Additionally, E11 noted that the activities could be easily implemented in the classroom: \textit{``yes, these activities (such as music appreciation) can be incorporated into upcoming lessons to increase student engagement.''}
I3 mentioned that the structured outcomes could potentially \textit{``make my teaching process more systematic.''}

\penguin{
We observed the different views between novice and expert teachers on how they embrace the integration of generated activities into real classrooms, though almost all teachers consider \name{}'s outcomes are beneficial for instructions. 
Experienced teachers (E9, E10, E11) preferred their self-centered teaching approach and were more conservative in using the generated activities. 
\textit{``To be honest, I have never used these (recommended) activities before, and in the future, I might only add a bit of them to my existing lesson plan to make students more interested.'' }(E11)
On the contrary, newer teachers were more open and tended to reconstruct their established curriculum based on the recommended activities.}
E8 stated, 
\textit{``I plan to adjust my instruction method according to them (activities); some recommended activities are excellent in the current context and can be adapted for use in other contexts.''
}

Despite these positive findings, five of nine teachers pointed out that some outcomes did not align well with the objectives of literature instruction: \textit{``the outcomes are still somewhat disconnected from practical application. The content generated based on the provided template does not fully meet the current teaching needs, especially considering the recently revised curriculum standards''} (E2).
Teachers suggested a potential solution: \textit{``Import the curriculum standards and teaching objectives for each text, and to consider these goals when constructing contexts. For instance, what are the learning objectives and abilities required for each grade level?''} (E5)

Despite modifications made to the prototype, 4 out of 9 experts indicated a need for additional outcomes that could directly enhance student learning in literature: \textit{``the documents generated are very useful for lesson preparation; however, they cannot be directly provided to students for learning purposes. I hope it can produce some homework questions''} (E1).

In summary, the comprehensive and structured outcomes facilitate educational activities; however, additional focus is required to ensure the alignment of these outcomes with the literature objectives.

\subsubsection{RQ2: Ideation process}
Experts have reached a consensus that the task load is low when using \name{} to construct contexts for the classroom, attributable to its well-structured layout and functional settings. E11 said, \textit{``the system does not burden my memory due to the collection feature''.} 
\penguin{
However, E5, who used \name{} in the wild for three days, suggested improvements to its UI design and features to further decrease mental demand: 
\textit{``More specific instructions could be incorporated into the page to clearly indicate the available actions. Additionally, I would like to have a feature that synchronizes generated records through user login for long-term usage.''} 
Experts in the lab sessions commented more on \name{}'s error cases which may increase their effort in the ideation process. 
For example, E8 input ``What activities can be designed around `Stepping Stones' [the title of one text]'', and \name{} responded ``Students can observe stone bridges and steps in their daily lives, explore their design principles and practical uses. They can participate in group projects to build stone bridge models using materials like rocks, experiencing the joy of collaboration.''
E8 commented, \textit{``This is indeed related to the article, but only the content, not the core idea or intended message''}.
E8 did not directly incorporate these outputs into the final outcome; instead, he edited manually.
Additionally, teachers may feel frustrated due to LLM's hallucinations.
I3 input 
``Give me more sentences and analysis related to 'Osmanthus Rain' [the title of one text] and the context about poetic life'', and
LLM responded 
```What I like is osmanthus. The osmanthus tree looks clumsy, unlike the plum tree, which has a graceful posture.' This contrast highlights the author's unique affection for osmanthus and reflects the author's ability to discover poetic elements in life through a lens of beauty''.
I3 remarked that \textit{``it did give me one more sentence but not really fit in the context.''}
In summary, \name{} is generally user-friendly but requires more UI and feature refinements to optimize the ideation process.
}

\subsubsection{RQ3: Perception of \name{}}
\penguin{
Both users in the lab session, who learned \name{} through verbal instructions from the developer, and users in the wild, who studied it via a video tutorial, agreed that our system is easy to learn.
}
With the exception of E2, all other teachers affirmed that \name{} supported their exploration of contexts and the interrelationships between texts and contexts. \textit{Expanding thinking} and \textit{inspiring creativity} were mentioned as advantages for supporting context exploration.
\textit{``The system provides a broader range of relationships between context and text, gradually expanding the context. I believe AI should work in this manner, incrementally broadening the scope to help me explore more possibilities, rather than rigidly offering a single answer''} (E11).
Furthermore, E5 added: \textit{``It can provide inspiration, demonstrating how to develop the class based on this context.''}
Besides expanding the breadth of thinking, E10 expressed appreciation for \name{}'s summarization capabilities: \textit{``after selecting a substantial number of texts, I was astonished that it could truly organize them and produce a comprehensive design, which is impossible in my typical lesson preparation.''} These comments indicate that \name{} facilitates users in opening and focusing their cognitive processes during interdisciplinary exploration.
\penguin{
However, E1 and E3 raised an issue about the repetition of suggested contexts when they tried different reading materials in their three-day usage of \name{}. 
E1 noted, \textit{``I found that the same contexts reappearing despite my selection of entirely different reading materials''.} 
}
Despite receiving feedback and modifications implemented after the prototype evaluation, this issue still troubled users and will be further discussed in~\ref{sec:discussion}.

In comparison to other generative tools and web search engines, users have reported a high level of trust in the output of our system, despite occasional acceptable errors. I3 stated, \textit{``I feel it is more closely aligned with reading materials compared to previous tools, although a few analyses of the context are not very accurate. Overall, I still trust this system.''} E4, I4 emphasized their reliance on personal experience and subjective judgment during the exploration process: \textit{``I always trust my own design more. When my ideas are limited, I use this system to evaluate its outputs and determine which content is usable''} (E4).

In summary, users perceive \name{} as highly usable and effective, and they generally express trust in its outputs.
\section{Discussion}\label{sec:discussion}
\subsection{Design Considerations \fanhx{for AI-Teacher Collaboration in Interdisciplinary Contexts}}
Based on the findings from two phases of experiments, we propose three design considerations for interdisciplinary context ideation.

\subsubsection{Encouraging Teacher Reflection Rather Than Full Automation}
One of the important features of \name{} is its high level of automation. During the exploration process, users can generate interdisciplinary contexts with detailed analyses and outcomes without the need for manual input unless asking the LLMs for answers.
This presents a significant advantage for experienced experts who possess a clear understanding of their requirements for classroom practice, as it alleviates their workload (\eg reducing the need for typing and searching), allowing them to concentrate fully on exploration.
However, findings from the within-subjects study indicated that novice teachers' behaviors suggest a potential risk associated with excessive automation. They might rely too heavily on the system, potentially neglecting their own cognitive process and allowing \name{} to dominate the exploration, which could result in content that is disproportionately focused on disciplinary or literature-specific outcomes.
\penguin{
Therefore, \name{} should encourage teachers to make reflections to improve their teaching. 
Such a focus is already central to other teaching tools, such as simulations that enhance student engagement~\cite{tang2024vizgroup} and help teachers understand students' learning states in online classrooms~\cite{ma2022glancee}.
}

\penguin{\subsubsection{Improving the Diversity and Suitability of Informal Educational Context Pool}
Based on DP4, we have collected 113 informal educational contexts from annual popular words in China and 144 subject-related contexts from textbooks to enable \name{} to suggest contexts.
}
In the within-subjects study, participants expressed appreciation for the diversity of contexts available in the laboratory setting. 
\penguin{
However, experts reported the lack of diverse and suitable contexts, especially the informal ones, when they would like to apply them in the literature courses. 
\fanhx{Besides, in the experiments, we observed 
\name{} retrieves similar contexts in its RAG process
to make recommendations.
While implementing advanced RAG strategies
might alleviate this,
we argue for the need for more diverse datasets.
}
The contexts sourced from Yao Wen Jiao Zi include popular words that Chinese teachers and students are likely to hear about, which could ease teachers' load in understanding the contexts. 
However, these popular words (\eg ``My youth is back!'', ``Prime Spot Debut'') may not always align seamlessly with pedagogical goals~\cite{sidekerskiene2024pedagogical} and may lean toward entertaining purposes. 
Therefore, future work could consider identifying suitable types of informal contexts with teachers and collecting more such contexts from diverse sources.
}
\fhx{
}

\subsubsection{Meeting the Expectations of Different Users}
\penguin{In accordance with DP1, we carefully designed \name{}'s workflow to align with teachers' practices and developed prompts using templates provided by E1-E7. 
However, our findings reveal that different user groups have varying expectations of the systems. 
For example, users without AI knowledge may need more expressive support that encourages them to criticize their own thoughts as well as the LLM output. 
Pre-service or new teachers could also be more depending on \name{} in the ideation process. 
\fanhx{To}
support and motivate specific users, 
\fanhx{collecting various templates from them and replace the original ones in \name{} can be beneficial (\eg prompting Text Reviewer to offer more incisive feedback can motivate teachers to think deeply).
Also, providing visualization of agent interactions and outputs can help users better understand the meaning of each step, to promote their critical thinking.}
}
\pzh{
\subsection{Reflections on AI-Teacher Collaboration}
}

\penguin{

Previous researchers on human-computer interaction and education have proposed various intelligent tutoring agents, \eg QuizBot \cite{ruan2019quizbot}, Sara \cite{winkler2020sara}, and DesignQuizzer \cite{peng2024designquizzer}. 
These agents act like teachers to prompt questions and give adaptive feedback to learners, which do supplement the lack of accessible teachers in online learning scenarios. 
Nevertheless, we argue that human educators are still irreplaceable, especially in offline classrooms. 
Teaching is not just about transferring knowledge but also about shaping students' cognitive, social, and emotional development~\cite{adeshola2023opportunities, chan2024will, guilherme2019ai}, which normally requires human teachers' interaction and empathy with students. 
Yet, these are qualities that AI currently lacks due to its generalization and lack of contextual awareness. 

Our design, development, and evaluation of \name{} provide a concrete example and implications of AI-teacher collaboration in such interdisciplinary teaching scenarios. 
In the design process, we identified with teachers the places where LLMs can help in their context ideation process, such as detailed analyses of the contexts, reading materials, and their relationships (DP2 \autoref{sec:principles}) as well as documented outcome (DP3). 
This human support helps AI improve its outcome lesson plans and promotion of knowledge transfer to that subject (\autoref{fig:rq1_results}). 
In addition, \name{} reduced users' workload in ideating contexts for literature teaching (\autoref{fig:rq2_results}), supporting that AI can serve as an auxiliary tool or a collaborator to improve the overall efficiency of teachers' tasks~\cite{holter2024deconstructing}. 
We implemented \name{} that aligns its LLM output as closely as possible with teachers' practices (DP1)
, but the novices' outcome lesson plans with \name{} did not align effectively with the teaching goals (\autoref{fig:rq1_results}), and experts also pointed out that some recommended contexts were unreasonable. 
It indicates that the iterative design of AI-teacher collaborative systems can be longitudinal, wherein the diverse teaching experiences can be utilized as data to refine AI-generated content through techniques such as fine-tuning or multi-modal fusion. 

We look forward to building an extensible version of \name{} in which teachers without programming skills can easily customize the system based on their needs to achieve more effective collaboration with AI. 
However, the ultimate responsibility for decisions of enacted teaching activities must remain with human educators. 
}

\penguin{
\subsection{Generalizability}
\fanhx{\name{} is open-source, allowing elementary school Chinese literature teachers to easily deploy and use it.} \name{} also has great potential to be generalized to support interdisciplinary teaching in elementary subjects apart from literature, in higher education, in other cultural contexts, and in self-learning tasks. 
For example, many elementary schools have explored STEM (science, technology, engineering, and mathematics) as an interdisciplinary approach to help students develop critical thinking, creativity, and problem-solving skills~\cite{english2015stem, sinatra2017speedometry}, and some universities have opened interdisciplinary programs in which students can take courses in various domains. 
\name{} can facilitate these educators with detailed analyses of the contexts and teaching materials, but to ensure it aligns with teachers' habitual practices, teachers' involvement in the design process is essential. 
To generalize \name{} and our findings to other cultural or language contexts, we should further be aware of the biases embedded in the context pool. 
Many of \name{}'s suggested contexts are sourced from annual popular words in China, which may potentially lead to misunderstanding or miscommunication in a multicultural environment. 
Lastly, while we position \name{} as a support tool for teachers, its analyses of reading materials and contexts could be useful for self-learners. 
For instance, learners can select the literature they are interested in, get recommended contexts from \name{}, and read them within a context in their after-school time. 
}

\subsection{Limitations and Future Work}
This study has several limitations. 
\penguin{
First, we only met with the head teacher E1 of the interdisciplinary literature teaching team (E1-E7) in the foundational study.
\fanhx{
It introduced limitations such as a one-sided perspective or inaccurate representation of information, which might challenge the robustness of the derived knowledge.
}
It would be more effective and insightful if we could directly reach them, \eg via a brainstorming session or design workshop, 
\fanhx{and involve more teaching teams}. 
}
Second, the participants in the within-subjects were students assigned to work as teachers, which means they do not possess adequate knowledge of teaching standards and the objectives of literature lessons.
Therefore, the data obtained may not accurately reflect the perspectives of actual teachers.
\fanhx{
Also, the within-subjects study design with a small sample size ($N=16$) significantly limits the robustness of our quantitative findings. 
Participants in our user study directly compare the systems rather than evaluate each system independently, which may lead to biased judgment.
Future studies should consider a between-subjects study design with a larger sample size.
}
\penguin{
Third, though we had an experienced teacher (E1) to score and comment on the outcome of ideation in Experiment I following the given criteria, it would further reduce the level of subjectivity involved in this rating task by inviting multiple experienced raters. 
}
\penguin{
Fourth, we evaluated \name{} in a short period and did not systematically study participants' usage patterns. 
A future field study is needed to observe teachers' usage of \name{} in their classrooms over a long period, \eg the adopted contexts and associated instructional methods they actually use.  
}
Fifth, our system is limited to textual content generation. Research indicates that the integration of multi-modal information (\eg images, and videos) can enhance educational activities. Future work could explore the incorporation of these elements into the ideation process within interdisciplinary contexts.
Finally, the algorithm implemented in \name{} is not yet the most advanced LLM technology. 
Incorporating more advanced RAG algorithms and multi-agent LLM architectures has the potential to enhance the quality of the system’s output.
\section{Conclusion}
In this paper, we developed an interactive system, \name{}, through an iterative design process involving 13 Chinese language teachers.
\name{} is to support the ideation of interdisciplinary contexts for teaching literature, with the help of LLM-generated content. \name{} includes three parts \ie Context Exploration View, Text Exploration View, and Collection.
It provides space that users can explore and curate contexts deemed suitable for literature teaching, and then find out the reading materials that can be integrated into the teaching process. 
\name{} provides detailed and precise analyses of the reading materials during exploration, subsequently providing structured outcomes for teachers. The within-subjects experiment involving 16 participants demonstrates that \name{} facilitate the depth of interdisciplinary integration and eases their workload during the exploration process.
Additionally, expert interviews were conducted with 9 teachers, who highlighted they trust in \name{} and believe it expands their thinking. We discuss design considerations and insights derived from the user study for human-AI collaboration in education.

\begin{acks}
This work is supported by the Young Scientists Fund of the National Natural Science Foundation of China with Grant No.: 62202509 and the General Projects Fund of the Natural Science Foundation of Guangdong Province in China with Grant No. 2024A1515012226.
\end{acks}

\bibliographystyle{ACM-Reference-Format}
\bibliography{references}

\appendix
\fanhx{
\section{Interview Questions in Expert Interviews} \label{sec:appendix}
}
The fixed questions for the \pzh{semi-structured} interviews in Experiment II (expert interviews) are shown below.

\penguin{
\textbf{RQ1: Ideation outcomes}}
\begin{itemize}
    \item What do you think about the quality of the outcomes generated by \name{}?
    \item Do you think the results generated by our system can assist you in preparing lessons or designing a new interdisciplinary context in the classroom setting?
\end{itemize}

\penguin{
\textbf{RQ2: Ideation process}}
\begin{itemize}
    \item Did you feel the task load is high while using \name{}? Specifically, did you feel any increased cognitive load or mental demands?
\end{itemize}

\penguin{
\textbf{RQ3: Perception of \name{}}
}
\begin{itemize}
    \item Do you think our system can assist you in ideating interdisciplinary topics more efficiently?
    \item Do you think our system can help you better explore connections between texts and the contexts?
    \item If you were to prepare an interdisciplinary reading case, compared to your usual preparation methods (\eg searching, meeting with other subject teachers, or using ChatGPT), do you think \name{} could support you better?
    \item Do you trust the output of our system compared to your previous experiences with web searching or using ChatGPT?
\end{itemize}



\end{document}